\documentclass[aps,pra,reprint,amsfonts,amsmath,amssymb]{revtex4-1}
\usepackage{graphicx}
\usepackage{dcolumn}
\usepackage{bm}
\usepackage{mathtools}
\usepackage{newtxtext}
\usepackage[varg]{newtxmath}
\usepackage[usenames]{color}
\usepackage[normalem]{ulem}
\usepackage{lipsum}

\DeclareMathOperator{\imaginary}{Im}
\DeclareMathOperator*{\Tr}{Tr}

\hypersetup{colorlinks=true,linkcolor=blue,citecolor=blue,urlcolor=blue}
\begin{document}

\title{Nonequilibrium susceptibility in photoinduced Floquet states}
\author{Atsushi Ono}
\author{Sumio Ishihara}
\affiliation{Department of Physics, Tohoku University, Sendai 980-8578, Japan}

\begin{abstract}
Nonequilibrium susceptibility in photoinduced Floquet states is studied.
We analyze an electron system coupled with a heat bath in a time-periodic oscillating electric field.
Spin/charge susceptibility is formulated on the basis of the Floquet Green function method, and is calculated numerically in a wide range of amplitude and frequency of light.
When the frequency is larger than the bandwidth, the susceptibility is enhanced due to the dynamical localization effect, and their peak positions in the momentum space are shifted by the Fermi surface deformation.
In the case of the small frequency and amplitude, multiple-peak structure emerges in the susceptibility, originating from the multiple Floquet bands which cross the Fermi level.
To confirm those numerical results and provide the interpretation, an approximated expression of the susceptibility is derived for small electric-field amplitude.
\end{abstract}
\maketitle

\section{Introduction}

Recent great progress in laser light technology promotes significantly the scientific research in strongly coupled light-matter systems.
Intense and ultrashort light irradiation induces a variety of nontrivial phenomena that are not realized in thermal equilibrium such as photoinduced phase transitions~\cite{Nasu2004,Tokura2006,Basov2011}, coherent control of spin or electronic polarization~\cite{Kirilyuk2010,Mentink2017,Kampfrath2013,Miyamoto2018}, and the dynamical localization (DL)~\cite{Kawakami2018}.
One of the fascinating phenomena induced by the strong light-matter coupling is appearance of the Floquet states, in which a time-periodic electromagnetic field behaves as photons and hybridizes with electrons~\cite{Shirley1965,Sambe1973,Aoki2014}.
In crystals, some replica energy bands termed Floquet bands are formed above and below the bare energy bands by photoirradiation.
The Floquet band structure, i.e., the one-particle excitation spectrum in the Floquet states, is described by the Floquet theory and experimentally detected through the time-resolved and angle-resolved photoemission spectroscopy~\cite{Wang2013,Mahmood2016}.
This has stimulated a number of studies on the ``Floquet engineering''~\cite{Bukov2015,Mentink2017,Eckardt2017,Oka2018}.

Besides the one-particle excitation spectra, the two-particle excitation channels are involved with electronic and structural properties as well as responses to external fields in nonequilibrium states~\cite{%
Eckstein2008,Tsuji2009,Tsuji2015,%
Perfetto2015b,Perfetto2015,Matsueda2007,Kanamori2011,Iyoda2014,Lenarcic2014,Kogoj2016,Shao2016,Shinjo2017,Ono2018,%
Bittner2018,%
Tsuji2016,Murakami2017,%
Fransson2010,Genkin1997,Fransson2010b,Power2012,Guimaraes2016,Stephanovich2017,Duan2018,Bauer2014,Ribeiro2015,Ribeiro2016,Ohnuma2017,Matsuo2018,%
Wang2018,%
Agarwalla2016,Golez2017,Ribeiro2017,%
Bunemann2017,%
Murakami2016,Murakami2016b,Fischer2018,Ido2017,%
Murakami2017b}.
The transient optical spectra observed by the pump-probe optical measurements are the representative examples, where the momentum transfer is limited to zero~\cite{Eckstein2008,Tsuji2009,Tsuji2015,Perfetto2015b,Perfetto2015,Matsueda2007,Kanamori2011,Iyoda2014,Lenarcic2014,Kogoj2016,Shao2016,Shinjo2017,Bittner2018,Tsuji2016,Murakami2017,Ono2018}.
The spin~\cite{Bittner2018,Fransson2010,Genkin1997,Fransson2010b,Power2012,Guimaraes2016,Stephanovich2017,Duan2018,Bauer2014,Ribeiro2015,Ribeiro2016,Ohnuma2017,Matsuo2018,Wang2018}, charge~\cite{Agarwalla2016,Golez2017,Ribeiro2017,Bunemann2017,Wang2018}, pair~\cite{Tsuji2016,Murakami2016,Ido2017,Murakami2016b,Murakami2017,Bunemann2017,Fischer2018,Wang2018}, and orbital~\cite{Murakami2017b} susceptibilities deduced from the two-particle excitation channels have also been investigated in a broad context, which are expected to diverge at a corresponding phase boundary of a photoinduced second-order phase transition.
The magnetic channel of the two-particle excitations governs exchange interactions, e.g., the Ruderman--Kittel--Kasuya--Yoshida (RKKY) interaction in magnetic metals and the superexchange interaction in Mott insulators.
Futhermore, the electron-hole excitations contribute to the lattice stability in electron-lattice coupled systems, known as the Peierls transition in low-dimensional materials.

In this paper, we examine the spin/charge susceptibility in photoinduced Floquet states.
We start with an electron system coupled with a heat bath in a continuous-wave (cw) electric field.
The susceptibility is formulated on the basis of the Floquet Green function method, and its approximated expression is derived from a series expansion.
The static susceptilities are numerically calculated in a wide range of the electric-field frequency ($\mathit{\Omega}$) and amplitude, as well as the electron number density.
In the case of the large $\mathit{\Omega}$ in comparison with the electron bandwidth, we found that the bandwidth reduction due to the DL effect enhances the susceptibility and deformation of the Fermi surface depending on the light polarization shifts the momenta at which susceptibilities take their maxima.
On the other hand, in the case of small $\mathit{\Omega}$ comparable to the bandwidth, the multiple Floquet bands cross the Fermi level and the additional peaks emerge at momenta corresponding to ``nesting vectors'' between the Floquet-band Fermi surfaces.

This paper is organized as follows.
In Sec.~\ref{sec:model}, we introduce the model Hamiltonian and the Floquet one-particle Green function.
Formulation of the susceptibility based on the Floquet Green function is presented in Sec.~\ref{sec:susceptibility}, and the expressions for weak electric-field amplitude are given in Sec.~\ref{sec:expansion}.
The numerical results in a two-dimensional square lattice are shown in Sec.~\ref{sec:2d}.
In Sec.~\ref{sec:1d}, the detailed analyses in a one-dimensional lattice are conducted.
Section~\ref{sec:summary} is devoted to summary.

\section{Formalism}
\subsection{Model and Floquet Green functions} \label{sec:model}
We consider a free-electron system coupled to a fermionic heat bath.
We adopt the Hamiltonian given by
\begin{align}
\mathcal{H} &= \mathcal{H}_{0}+\mathcal{H}', \\
\mathcal{H}_{0} &= \sum_{\bm{k}s} \varepsilon_{\bm{k}} c_{\bm{k}s}^\dagger c_{\bm{k}s} + \sum_{\nu} \varepsilon_{\nu} b_{\nu}^\dagger b_{\nu}, \\
\mathcal{H}' &= \sum_{\bm{k}s\nu} V_{\nu} {\left( c_{\bm{k}s}^\dagger b_{\nu} + b_{\nu}^\dagger c_{\bm{k}s} \right)},
\end{align}
where $c_{\bm{k}s}^\dagger$ is a creation operator of an electron with momentum $\bm{k}$ and spin $s\ (={\uparrow},{\downarrow})$, and $b_{\nu}^\dagger$ is that of a fermion in a bath with quantum number $\nu$.
The first term $\mathcal{H}_0$ describes the free-electron system and the bath, and the second term $\mathcal{H}'$ represents the coupling between them.
The electron energy band (measured from the chemical potential) and the bath energy level are denoted by $\varepsilon_{\bm{k}}$ and $\varepsilon_\nu$, respectively.
The hybridization matrix element $V_{\nu}$ is independent of $\bm{k}$, implying the local coupling.
A vector potential of a cw field, $\bm{A}(t)$, at time $t$ is incorporated in $\varepsilon_{\bm{k}}$ as the Peierls phase as $\varepsilon_{\bm{k}} \mapsto \varepsilon_{\bm{k}-e\bm{A}(t)/\hbar}$, where $e\ (<0)$ is the electron charge, and $\hbar$ is the reduced Planck constant.
We consider both the linearly polarized light defined by
\begin{align}
\bm{A}(t) &= \bm{A}_0 \sin \mathit{\Omega} t = (A_0^x \sin \mathit{\Omega} t, A_0^y \sin \mathit{\Omega} t) ,
\label{eq:def_linear_pol}
\end{align}
and the circularly polarized light defined by
\begin{align}
\bm{A}(t) &= A_0 (\cos \mathit{\Omega} t, \sin \mathit{\Omega} t),
\label{eq:def_circ_pol}
\end{align}
where $A_0=\Vert \bm{A}_0 \Vert = \sqrt{(A_0^x)^2+(A_0^y)^2}$ and $\mathit{\Omega}$ are amplitude and frequency of the vector potential, respectively, in the two-dimensional square lattice.
In the case of the one-dimensional lattice, we define $A(t) = A_0 \sin \mathit{\Omega} t$.
We suppose that the initial state of the electron system before irradiation is a paramagnetic metallic state:
\begin{align}
\vert \Psi_0 \rangle = \prod_{\bm{k}s}^{\varepsilon_{\bm{k}}<0} c_{\bm{k}s}^\dagger \vert 0 \rangle,
\end{align}
where $\vert 0 \rangle$ is a vacuum of the electrons.
From now on, the reduced Planck constant $\hbar$, the electron charge $e$, and the lattice constant are taken to be one.

We introduce the Floquet Green function and a bath selfenergy (see, e.g., Ref.~\cite{Aoki2014} for details).
We define the retarded, advanced, and lesser Green functions of the electrons as
\begin{align}
G_{\bm{k}}^{\mathrm{R}}(t,t')
&= -i\theta(t-t') \langle \{c_{\bm{k}s}(t), c_{\bm{k}s}^\dagger(t')\} \rangle,
\label{eq:green_retarded_twotime} \\
G_{\bm{k}}^{\mathrm{A}}(t,t')
&= G_{\bm{k}}^{\mathrm{R}}(t',t)^*, \\
G_{\bm{k}}^{<}(t,t')
&= i\langle c_{\bm{k}s}^\dagger(t') c_{\bm{k}s}(t) \rangle,
\label{eq:green_lesser_twotime}
\end{align}
respectively, where $\theta(t)$ is the step function, $\{\cdot ,\cdot \}$ denotes the anticommutator, and $\langle {\cdot} \rangle = \langle \Psi_0 \vert {\cdot} \vert \Psi_0 \rangle$ represents the expectation value.
Here, the operators are given in the Heisenberg picture, where the time-evolution is governed by the full Hamiltonian $\mathcal{H}$.
Since we focus on the paramagnetic state, the Green functions are independent of the spin $s$,
and the spin indices in the left hand sides in Eqs.~\eqref{eq:green_retarded_twotime}--\eqref{eq:green_lesser_twotime} are omitted.
In the steady states driven by the cw field, the two-time Green functions defined above have the following time periodicity:
\begin{align}
G^X(t+T,t'+T) = G^X(t,t') ,
\end{align}
where $X= \mathrm{R},\mathrm{A},{<}$ and $T=2\pi/\mathit{\Omega}$.
This periodicity enables one to introduce the Floquet representation, called the Floquet Green function, as
\begin{align}
(G^X)_{mn}(\omega)
&= \int_0^T \frac{dt_a}{T} \int_{-\infty}^{\infty} dt_r\, e^{i(\omega+m\mathit{\Omega})t-i(\omega+n\mathit{\Omega})t'} G^X(t,t'),
\label{eq:floquet_twotime}
\end{align}
where $t_a=(t+t')/2$ and $t_r=t-t'$.
The indices $m$ and $n$ take integers, which are restricted to $\{ 0,\pm 1,\pm 2,\dots, \pm N_p \}$ in the present numerical calculations.
The inverse transformation of Eq.~\eqref{eq:floquet_twotime} is defined by
\begin{align}
G^X(t,t')
&= \sum_n \int_{-\infty}^{\infty} \frac{d\omega}{2\pi}\, e^{-in\mathit{\Omega} t_a} e^{-i(\omega+(n/2)\mathit{\Omega})t_r} (G^X)_{n,0}(\omega).
\label{eq:twotime_floquet}
\end{align}
Equations~\eqref{eq:floquet_twotime} and \eqref{eq:twotime_floquet} are applied to any two-time functions with the same periodicity.
The retarded Floquet Green function $(G_{\bm{k}}^{\mathrm{R}})_{mn}(\omega)$ is obtained from the Dyson equation given by
\begin{align}
(G_{\bm{k}}^{\mathrm{R},-1})_{mn}(\omega)
&= (\mathcal{G}_{\bm{k}}^{\mathrm{R},-1})_{mn}(\omega) - (\mathit{\Sigma}_{\bm{k}}^{\mathrm{R}})_{mn}(\omega) .
\label{eq:dyson_retarded}
\end{align}
Here, $\mathit{\Sigma}_{\bm{k}}^{\mathrm{R}}$ is the retarded selfenergy and $\mathcal{G}_{\bm{k}}^{\mathrm{R}}$ is the bare Green function:
\begin{align}
(\mathcal{G}_{\bm{k}}^{\mathrm{R},-1})_{mn}(\omega)
&= \delta_{mn}(\omega+n\mathit{\Omega}+i\eta) - \varepsilon_{m-n,\bm{k}},
\end{align}
where $\eta$ is a positive infinitesimal and $\varepsilon_{n,\bm{k}}$ is the $n$th Fourier component of $\varepsilon_{\bm{k}-\bm{A}(t)}$ defined by
\begin{align}
\varepsilon_{n,\bm{k}} = \int_0^T \frac{dt}{T}\, e^{in\mathit{\Omega} t} \varepsilon_{\bm{k}-\bm{A}(t)}.
\label{eq:epsilon_floquet}
\end{align}
In particular, $\varepsilon_{0,\bm{k}}$ is a time average of $\varepsilon_{\bm{k}-\bm{A}(t)}$ during the time period and is reduced to $\varepsilon_{\bm{k}}$ at $A_0=0$.
We obtain the selfenergy by integrating out the bath degrees of freedom as
\begin{align}
(\mathit{\Sigma}_{\bm{k}}^{X})_{mn}(\omega)
&= \delta_{mn} \sum_{\nu} \vert V_\nu \vert^2 \mathcal{F}_{\nu}^{X}(\omega+n\mathit{\Omega}),
\end{align}
where $\mathcal{F}_\nu^X(\omega)$ is the bare Green function of the bath in the Wigner representation, i.e., the Fourier transformation of the two-time Green function $\mathcal{F}_\nu^X(t,t')$ with respect to $t_r=t-t'$.
These are defined by
\begin{align}
\mathcal{F}_\nu^{\mathrm{R}}(\omega) &= \mathcal{F}_\nu^{\mathrm{A}}(\omega)^* = \frac{1}{\omega-\varepsilon_\nu+i\eta}, \\
\mathcal{F}_\nu^{<}(\omega) &= 2\pi i f(\omega) \delta(\omega-\varepsilon_\nu),
\end{align}
where $\delta(\omega)$ is the Dirac delta function and $f(\omega)=1/(e^{\beta \omega}+1)$ is the Fermi--Dirac function with the inverse temperature of the bath, $\beta$.
For simplicity, we assume that the energy spectrum of the bath is broad enough that the real parts of the retarded and advanced selfenergies are included into the electron chemical potential, and the imaginary parts of them are independent of $\omega$.
This leads to
\begin{align}
(\mathit{\Sigma}_{\bm{k}}^{\mathrm{R}})_{mn}(\omega) 
&\approx -\delta_{mn} i \mathit{\Gamma}, \\
(\mathit{\Sigma}_{\bm{k}}^{<})_{mn}(\omega) 
&\approx 2 \delta_{mn} i \mathit{\Gamma} f(\omega+n\mathit{\Omega}),
\end{align}
where $\mathit{\Gamma} \equiv \pi \sum_\nu \vert V_\nu \vert^2 \delta(\omega-\varepsilon_\nu) \ (>0)$ represents the coupling strength between the system and the bath.
The Dyson equation in Eq.~\eqref{eq:dyson_retarded} is now written as
\begin{align}
(G_{\bm{k}}^{\mathrm{R},-1})_{mn}(\omega)
&= \delta_{mn}(\omega+n\mathit{\Omega}+i\mathit{\Gamma})-\varepsilon_{m-n,\bm{k}},
\label{eq:dyson_retarded_2}
\end{align}
where the positive infinitesimal $\eta$ is replaced by the coupling strength $\mathit{\Gamma}$.
According to Ref.~\cite{Tsuji2008}, the retarded Floquet Green function, i.e., the inverse of Eq.~\eqref{eq:dyson_retarded_2}, is given by
\begin{align}
(G_{\bm{k}}^{\mathrm{R}})_{mn}(\omega)
&= \sum_l \frac{(\mathit{\Lambda}_{\bm{k}})_{ml} (\mathit{\Lambda}_{\bm{k}})_{nl}^*}{\omega+l\mathit{\Omega}-\varepsilon_{0,\bm{k}}+i\mathit{\Gamma}},
\label{eq:green_retarded_floquet}
\end{align}
where $(\mathit{\Lambda}_{\bm{k}})_{mn}$ is the unitary matrix defined by
\begin{align}
(\mathit{\Lambda}_{\bm{k}})_{mn}
&= \int_{-\pi}^{\pi} \frac{dx}{2\pi}\, e^{i(m-n)x} \notag \\
&\quad \times \exp {\left[ \frac{1}{i\mathit{\Omega}}
\int_0^x dz\,
{\left\{ \varepsilon_{\bm{k}-\bm{A}(z/\mathit{\Omega})}-\varepsilon_{0,\bm{k}} \right\}} \right]}.
\label{eq:def_lambda}
\end{align}
Equation~\eqref{eq:green_retarded_floquet} indicates that, in the steady states, ``$l$-photon-dressed'' sidebands with energy $\varepsilon_{0,\bm{k}}-l\mathit{\Omega}$ emerge around the ``zero-photon'' band $\varepsilon_{0,\bm{k}}$.
The advanced and lesser Floquet Green functions are obtained as
\begin{align}
(G_{\bm{k}}^{\mathrm{A}})_{mn}(\omega)
&= (G_{\bm{k}}^{\mathrm{R}})_{nm}(\omega)^*, \label{eq:green_advanced_floquet} \\
(G_{\bm{k}}^{<})_{mn}(\omega)
&= ( G_{\bm{k}}^{\mathrm{R}} \mathit{\Sigma}_{\bm{k}}^{<} G_{\bm{k}}^{\mathrm{A}} )_{mn}(\omega) ,
\label{eq:green_lesser_floquet}
\end{align}
respectively.
We note that the unitary matrix $(\mathit{\Lambda}_{\bm{k}})_{mn}$ is reduced to the identity: $(\mathit{\Lambda}_{\bm{k}})_{mn}=\delta_{mn}$ in the case of $A_0=0$ or in the limit of $\mathit{\Omega}\rightarrow\infty$, where the retarded and lesser Green functions are written as
\begin{align}
(G_{\bm{k}}^{\mathrm{R}})_{mn}(\omega)
&= \frac{\delta_{mn}}{\omega+n\mathit{\Omega}-\varepsilon_{0,\bm{k}}+i\mathit{\Gamma}},
\label{eq:green_retarded_floquet_eq} \\
(G_{\bm{k}}^{<})_{mn}(\omega)
&= \frac{2 \delta_{mn} i \mathit{\Gamma} f(\omega+n\mathit{\Omega})}{(\omega+n\mathit{\Omega}-\varepsilon_{0,\bm{k}})^2+\mathit{\Gamma}^2}.
\label{eq:green_lesser_floquet_eq}
\end{align}

On the other hand, the unitary matrix in Eq.~\eqref{eq:def_lambda} is singular in the low-frequency limit ($\mathit{\Omega} \rightarrow 0$)~\footnote{In this paper, we call the limits of $\mathit{\Omega}\rightarrow 0$ and $\omega\rightarrow 0$ the low-frequency limit and the static limit, respectively}.
Thus, one has to go back to Eq.~\eqref{eq:dyson_retarded_2}, which reads
\begin{align}
(G_{\bm{k}}^{\mathrm{R},-1})_{mn}(\omega)
= \delta_{mn}(\omega+i\mathit{\Gamma})-\varepsilon_{m-n,\bm{k}}.
\end{align}
This matrix in the Floquet space is analogous to a bilinear Hamiltonian of the one-dimensional tight-binding model where the ``on-site potential'' is $\omega+i\mathit{\Gamma}$ and the hopping amplitude between the $m$th and $n$th ``sites'' is $\varepsilon_{m-n,\bm{k}}$.
The $(2N_p+1)$-dimensional matrix $\varepsilon_{mn,\bm{k}}\equiv \varepsilon_{m-n,\bm{k}}$ in Eq.~\eqref{eq:epsilon_floquet} is diagonalized by the Fourier transformation associated with the unitary matrix $U_{n\kappa} = e^{-in\kappa}/\sqrt{2N_p+1}$ with the ``wavenumber'' $\kappa = 2\pi j/(2N_p+1) \ (j=0,\pm 1,\dots, \pm N_p)$.
The eigenvalue of $\varepsilon_{mn,\bm{k}}$ is given by
\begin{align}
\tilde{\varepsilon}_{\kappa,\bm{k}}
&= \sum_{mn} U_{m\kappa}^* \varepsilon_{mn,\bm{k}} U_{n\kappa} \notag \\
&= \sum_{n} \varepsilon_{n,\bm{k}} e^{in\kappa}
= \varepsilon_{\bm{k}-\bm{A}(-\kappa/\mathit{\Omega})},
\end{align}
where the energy band $\varepsilon_{\bm{k}}$ is shifted by $\bm{A}(-\kappa/\mathit{\Omega})$ in the momentum space with $\kappa/\mathit{\Omega}$ corresponding to time.
Then, the Floquet Green function is written as
\begin{align}
(G_{\bm{k}}^{\mathrm{R}})_{mn}(\omega)
&= \frac{1}{2N_p+1} \sum_{\kappa} \frac{e^{-i(m-n)\kappa}}{\omega+i\mathit{\Gamma}-\tilde{\varepsilon}_{\kappa,\bm{k}}}
\label{eq:green_retarded_lowfreq} \\
&\rightarrow \int_{-T/2}^{T/2} \frac{dt}{T} \frac{e^{i(m-n)\mathit{\Omega} t}}{\omega+i\mathit{\Gamma}-\varepsilon_{\bm{k}-\bm{A}(t)}},
\label{eq:green_retarded_lowfreq_continuous} \\
(G_{\bm{k}}^{<})_{mn}(\omega)
&= \frac{1}{2N_p+1} \sum_{\kappa} \frac{2i \mathit{\Gamma} f(\omega) e^{-i(m-n)\kappa}}{(\omega-\tilde{\varepsilon}_{\kappa,\bm{k}})^2+\mathit{\Gamma}^2}
\label{eq:green_lesser_lowfreq} \\
&\rightarrow \int_{-T/2}^{T/2} \frac{dt}{T} \frac{2i\mathit{\Gamma} f(\omega) e^{i(m-n)\mathit{\Omega} t}}{(\omega-\varepsilon_{\bm{k}-\bm{A}(t)})^2+\mathit{\Gamma}^2},
\label{eq:green_lesser_lowfreq_continuous}
\end{align}
where we take the limit of $N_p\rightarrow \infty$ in Eqs.~\eqref{eq:green_retarded_lowfreq_continuous} and \eqref{eq:green_lesser_lowfreq_continuous}.
These are the $(m-n)$th Fourier components of the equilibrium Green functions in which $\varepsilon_{\bm{k}}$ is replaced by $\varepsilon_{\bm{k}-\bm{A}(t)}$.
We evaluate Eq.~\eqref{eq:green_retarded_lowfreq} in the two-dimensional square lattice, which is shown in Fig.~\ref{fig:etadep}(j) in Sec.~\ref{sec:2d}.

We define the spectral functions as the imaginary parts of the time-averaged Green functions:
\begin{align}
\rho^{\mathrm{R}}_{\bm{k}}(\omega) &= -\frac{1}{\pi} \imaginary {(G_{\bm{k}}^{\mathrm{R}})_{00}(\omega)}, \\
\rho^{<}_{\bm{k}}(\omega) &= \frac{1}{2\pi} \imaginary {(G_{\bm{k}}^{<})_{00}(\omega)}.
\end{align}
The density of states and the momentum distribution function are given by
\begin{align}
n(\omega) &= \frac{2}{N} \sum_{\bm{k}} \rho_{\bm{k}}^{\mathrm{R}}(\omega), \\
n_{\bm{k}} &= \int_{-\infty}^{\infty} d\omega\, \rho_{\bm{k}}^{<}(\omega),
\end{align}
respectively, where the prefactor $2$ in $n(\omega)$ reflects the spin degree of freedom.

\subsection{Spin and charge susceptibilities} \label{sec:susceptibility}
We consider the spin and charge densities with wavenumber $\bm{q}$ defined by
\begin{align}
M_{\bm{q}}^\alpha = \frac{1}{N} \sum_{\bm{k}ss'} \sigma_{ss'}^\alpha c_{\bm{k}s}^\dagger c_{\bm{k}+\bm{q}s'} ,
\end{align}
for $\alpha \in \{0,1,2,3\}$, where $\sigma^0$ is the identity matrix and $\{ \sigma^1, \sigma^2, \sigma^3 \}$ are the Pauli matrices, and $N$ denotes the number of the lattice sites.
We introduce the four-vector notation: $M_{\bm{q}}^{\alpha}=(M_{\bm{q}}^0,\bm{M}_{\bm{q}})$ with $\bm{M}_{\bm{q}}=(M_{\bm{q}}^1,M_{\bm{q}}^2,M_{\bm{q}}^3)$.
Hamiltonian for the coupling between $M_{\bm{q}}^{\alpha}$ and an external field $H_{\bm{q}}^{\alpha}$ (i.e., a scalar potential for $\alpha=0$ and a magnetic field for $\alpha=1,2,3$) is given by
\begin{align}
\mathcal{H}_{\text{ext}} = - \sum_{\alpha\bm{q}} H_{\bm{q}}^{\alpha} M_{\bm{q}}^{\alpha}
= - \sum_{\mathclap{\alpha\bm{k}\bm{q}ss'}} H_{\bm{q}}^{\alpha} \sigma_{ss'}^{\alpha} c_{\bm{k}s}^\dagger c_{\bm{k}+\bm{q},s'}.
\end{align}
The spin and charge susceptibilities are defined by the functional derivative:
\begin{align}
\chi_{\bm{q}\bm{q}'}^{\alpha\beta}(t,t')
= \frac{\delta \langle M_{\bm{q}}^{\alpha}(t) \rangle}{\delta H_{\bm{q}'}^{\beta}(t')}.
\end{align}
Following Ref.~\cite{Ono2018} and references therein, we obtain
\begin{widetext}
\begin{align}
\chi_{\bm{q}\bm{q}'}^{\alpha\beta}(t,t')
= \frac{i\delta_{\bm{q}\bm{q}'}}{N} \sum_{\bm{k}} \sum_{ss'} {\left[ \sigma_{ss'}^\alpha G_{\bm{k}+\bm{q}}^{\mathrm{R}}(t,t') \sigma_{s's}^\beta G_{\bm{k}}^{<}(t',t) + \sigma_{ss'}^\alpha G_{\bm{k}+\bm{q}}^{<}(t,t') \sigma_{s's}^\beta G_{\bm{k}}^{\mathrm{A}}(t',t) \right]},
\label{eq:susceptibility_twotime}
\end{align}
where we assume that the system is in the homogeneous paramagnetic state.
This satisfies the causality since $G^{\mathrm{R}}(t,t')$ and $G^{\mathrm{A}}(t',t)$ are proportional to $\theta(t-t')$.
The susceptibility is written in the Floquet representation as
\begin{align}
(\chi_{\bm{q}})_{mn}(\omega)
= \frac{2i}{N} \sum_{\bm{k}l} \int_{-\infty}^{\infty} \frac{d\omega'}{2\pi} {\biggl[
(G_{\bm{k}+\bm{q}}^{\mathrm{R}})_{m,n+l}(\omega+\omega')
(G_{\bm{k}}^{<})_{l,0}(\omega')
+ (G_{\bm{k}+\bm{q}}^{<})_{m,n+l}(\omega+\omega')
(G_{\bm{k}}^{\mathrm{A}})_{l,0}(\omega')
\biggr]},
\label{eq:susceptibility_floquet}
\end{align}
\end{widetext}
where we use $\Tr (\sigma^\alpha \sigma^\beta) = 2\delta_{\alpha\beta}$, and omit the indices $\alpha$, $\beta$, and $\bm{q}'$ by taking $\beta=\alpha$ and $\bm{q}'=\bm{q}$.
The susceptibility in Eq.~\eqref{eq:susceptibility_floquet} is independent of $\alpha$, which means that the magnetic susceptibility is isotropic in spin space and coincides with the charge susceptibility in the present system.
We focus on the time average of the susceptibility,
\begin{align}
\chi_{\bm{q}}(\omega) \equiv (\chi_{\bm{q}})_{nn}(\omega-n\mathit{\Omega}),
\label{eq:susceptibility_dynamical}
\end{align}
and its static limit ($\omega\rightarrow 0$),
\begin{align}
\chi_{\bm{q}} \equiv \chi_{\bm{q}}(0)
= (\chi_{\bm{q}})_{nn}(-n\mathit{\Omega}).
\label{eq:susceptibility_static}
\end{align}
As mentioned in Sec.~\ref{sec:model}, the unitary matrix in Eq.~\eqref{eq:def_lambda} is reduced to $(\mathit{\Lambda}_{\bm{k}})_{mn}=\delta_{mn}$ in the case of $A_0=0$ or in the limit of $\mathit{\Omega}\rightarrow \infty$, where the Green functions are given in Eqs.~\eqref{eq:green_retarded_floquet_eq} and \eqref{eq:green_lesser_floquet_eq}.
This simplifies the susceptibility in Eq.~\eqref{eq:susceptibility_floquet} to the following form:
\begin{align}
\chi_{\bm{q}}(\omega)
&\rightarrow \chi_{\bm{q}}^{(0)}(\omega)
\equiv \frac{2}{N} \sum_{\bm{k}} \frac{f(\varepsilon_{0,\bm{k}+\bm{q}})-f(\varepsilon_{0,\bm{k}})}{\omega-(\varepsilon_{0,\bm{k}+\bm{q}}-\varepsilon_{0,\bm{k}})+2i\eta},
\label{eq:susceptibility_highfreq}
\end{align}
where we take the limit of $\mathit{\Gamma}\rightarrow \eta$ in order to replace $f(\omega)$ with $f(\varepsilon_{0,\bm{k}})$ or $f(\varepsilon_{0,\bm{k}+\bm{q}})$.

In the low-frequency limit ($\mathit{\Omega}\rightarrow 0$), where the Green functions are given in Eqs.~\eqref{eq:green_retarded_lowfreq} and \eqref{eq:green_lesser_lowfreq}, the time-averaged susceptibility in Eq.~\eqref{eq:susceptibility_floquet} is evaluated as
\begin{align}
(\chi_{\bm{q}})_{00}(\omega)
&\rightarrow \frac{2}{(2N_p+1)N} \sum_{\bm{k}\kappa} \frac{f(\tilde{\varepsilon}_{\kappa,\bm{k}+\bm{q}})-f(\tilde{\varepsilon}_{\kappa,\bm{k}})}{\omega-(\tilde{\varepsilon}_{\kappa,\bm{k}+\bm{q}}-\tilde{\varepsilon}_{\kappa,\bm{k}})+2i\eta} \notag \\
&= \frac{2}{N} \sum_{\bm{k}} \frac{f(\varepsilon_{\bm{k}+\bm{q}})-f(\varepsilon_{\bm{k}})}{\omega-(\varepsilon_{\bm{k}+\bm{q}}-\varepsilon_{\bm{k}})+2i\eta},
\label{eq:susceptibility_lowfreq}
\end{align}
which is the well-known formula for the susceptibility in equilibrium systems.
This is interpreted as follows:
the typical timescale for the system to reach the steady state is given by $\mathit{\Gamma}^{-1}$, while $(\chi_{\bm{q}})_{00}(\omega)$ represents the susceptibility averaged during a time interval $[0,T{=}2\pi \mathit{\Omega}^{-1}]$.
We note that the low-frequency cw field is essentially different from the static external field, since we consider the time-averaged susceptibility and Green functions in the Floquet representation;
once the external field is applied along a certain direction during $t\in[0,T/2)$, then the external field is inevitably applied in the opposite direction during $t\in[T/2,T)$.
In the low-frequency limit ($\mathit{\Omega} \ll \mathit{\Gamma}$), the system is considered to be always in the equilibrated state where the energy band is shifted by $\bm{A}(t)$ in the momentum space and the electron distribution function is given by the Fermi--Dirac function.
Therefore, we end up with the well-known expression of the equilibrium susceptibility in Eq.~\eqref{eq:susceptibility_lowfreq}.
This picture is numerically confirmed for the case of the finite but small frequency $\mathit{\Omega}$ in Sec.~\ref{sec:2d} (see Fig.~\ref{fig:etadep}).

Here, we mention the relation of the susceptibility in Eq.~\eqref{eq:susceptibility_floquet} to the RKKY interaction~\cite{Ruderman1954,Kasuya1956,Yosida1957}.
Let us suppose that two magnetic impurities described by classical spins $\bm{S}_i\ (i=1,2)$ are immersed in the conduction sea at the positions $\bm{r}_i$.
These spins couple to the spin density of the conduction electrons, $\bm{M}_i$, represented by Hamiltonian $-J\sum_{i} \bm{S}_i \bm{M}_i$, where we define $\bm{M}_i=\sum_{ss'} \bm{\sigma}_{ss'} c_{is}^\dagger c_{is'}$ with $c_{is}^\dagger = N^{-1/2} \sum_{\bm{k}} e^{-i\bm{k}\bm{r}_i} c_{\bm{k}s}^\dagger$, and $J$ is a coupling constant.
The conduction electron at $\bm{r}_j$ feels a magnetic field $\bm{H}_j=J\bm{S}_j$, which induces the spin density at $\bm{r}_i$ as $\langle M_i^\alpha(t) \rangle \equiv \sum_{\beta} \int_{-\infty}^{\infty} dt'\, \chi_{ij}^{\alpha\beta}(t,t') H_j^{\beta}(t')$.
Thus, the magnetic interaction between the impurities mediated by the conduction electrons, i.e., the RKKY interaction, is given by
\begin{align}
\mathcal{H}_{\text{RKKY}}(t)
&=-J\bm{S}_i(t) \langle \bm{M}_i(t) \rangle \notag \\
&= -J^2 \sum_{\alpha\beta} \int_{-\infty}^{\infty} dt'\, S_{i}^{\alpha}(t) \chi_{ij}^{\alpha\beta}(t,t') S_j^\beta(t'),
\label{eq:rkky_twotime}
\end{align}
where $\chi_{ij}^{\alpha\beta}(t,t')$ is given from $\chi_{\bm{q}}^{\alpha\beta}(t,t')$ in Eq.~\eqref{eq:susceptibility_twotime} as
\begin{align}
\chi_{ij}^{\alpha\beta}(t,t')
= \frac{1}{N} \sum_{\bm{q}} e^{i\bm{q}(\bm{r}_i-\bm{r}_j)} \chi_{\bm{q}}^{\alpha\beta}(t,t').
\label{eq:susceptibility_realspace1}
\end{align}
When the timescales of the impurity dynamics are much slower than $\mathit{\Omega}^{-1}$, the time dependence of $\bm{S}_i(t)$ can be neglected.
In this case, the interaction in Eq.~\eqref{eq:rkky_twotime} is written as
\begin{align}
\mathcal{H}_{\text{RKKY}}(t)
= -J^2 \sum_{\alpha\beta} S_i^\alpha S_j^\beta \sum_n e^{-in\mathit{\Omega} t} (\chi_{ij}^{\alpha\beta})_{n,0}(0),
\end{align}
and its time average over the interval $[0,T{=}2\pi \mathit{\Omega}^{-1}]$ takes the following form:
\begin{align}
\int_0^T \frac{dt}{T}\, \mathcal{H}_{\text{RKKY}}(t)
= -J^2 \sum_{\alpha\beta} S_i^\alpha (\chi_{ij}^{\alpha\beta})_{00}(0) S_i^\beta.
\end{align}
This result implies that the RKKY interaction in the cw field can be estimated by the static susceptibility given in Eq.~\eqref{eq:susceptibility_floquet}, as in the case of the equilibrium states.

\subsection{Series expansion of the susceptibility}
\label{sec:expansion}

In this section, we show results of a series expansion of the susceptibility in Eq.~\eqref{eq:susceptibility_floquet} with respect to the vector-potential amplitude $A_0$.
We consider the linearly polarized light in Eq.~\eqref{eq:def_linear_pol}.
The details of the derivation are presented in Appendix.

The unitary matrix $(\mathit{\Lambda}_{\bm{k}})_{mn}$ in Eq.~\eqref{eq:def_lambda} is expanded as
\begin{widetext}
\begin{align}
(\mathit{\Lambda}_{\bm{k}})_{mn}
&= \delta_{mn} - \frac{\bm{v}_{\bm{k}}\bm{A}_0}{i\mathit{\Omega}} {\left( \delta_{mn} - \frac{\delta_{m,n-1}+\delta_{m,n+1}}{2} \right)}
+ {\left(\frac{\bm{v}_{\bm{k}}\bm{A}_0}{i\mathit{\Omega}}\right)}^2 {\left( \frac{3\delta_{mn}}{4} - \frac{\delta_{m,n-1}+\delta_{m,n+1}}{2} + \frac{\delta_{m+1,n-1}+\delta_{m-1,n+1}}{8} \right)} \notag \\
&\quad + \sum_{\alpha\beta} \frac{\tau_{\bm{k}}^{\alpha\beta}A_0^\alpha A_0^\beta}{16\mathit{\Omega}} (\delta_{m+1,n-1}-\delta_{m-1,n+1})
+ \mathcal{O}(A_0^3),
\label{eq:expansion_lambda}
\end{align}
where $\bm{v}_{\bm{k}} = \partial\varepsilon_{\bm{k}}/\partial \bm{k}$ is the group velocity and $\tau_{\bm{k}}^{\alpha\beta} = \partial^2 \varepsilon_{\bm{k}}/\partial k^\alpha \partial k^\beta$ is the energy stress tensor.
The indices $\alpha$ and $\beta$ run over $\{1,2,3\}$.
The time-averaged component of the retarded and lesser Green functions are written as
\begin{align}
(G_{\bm{k}}^{\mathrm{R}})_{00}(\omega)
&= {\left( 1-\frac{\mathcal{A}_{\bm{k}}^2}{2} \right)} \frac{1}{\omega-\varepsilon_{0,\bm{k}}+i\eta}
+ \frac{\mathcal{A}_{\bm{k}}^2}{4} \frac{1}{\omega+\mathit{\Omega}-\varepsilon_{0,\bm{k}}+i\eta}
+ \frac{\mathcal{A}_{\bm{k}}^2}{4} \frac{1}{\omega-\mathit{\Omega}-\varepsilon_{0,\bm{k}}+i\eta}
+ \mathcal{O}(A_0^3),
\label{eq:expansion_retarded_00} \\
\frac{(G_{\bm{k}}^{<})_{00}(\omega)}{2i\eta}
&= \frac{(1-\mathcal{A}_{\bm{k}}^2) f(\omega)}{(\omega-\varepsilon_{0,\bm{k}})^2+\eta^2}
+ \frac{\mathcal{A}_{\bm{k}}^2}{4} \frac{f(\omega+\mathit{\Omega})+f(\omega-\mathit{\Omega})}{(\omega-\varepsilon_{0,\bm{k}})^2+\eta^2}
+ \frac{\mathcal{A}_{\bm{k}}^2}{4} {\left[
\frac{f(\omega+\mathit{\Omega})}{(\omega+\mathit{\Omega}-\varepsilon_{0,\bm{k}})^2+\eta^2}
+ \frac{f(\omega-\mathit{\Omega})}{(\omega-\mathit{\Omega}-\varepsilon_{0,\bm{k}})^2+\eta^2}
\right]} 
+ \mathcal{O}(A_0^3),
\label{eq:expansion_lesser_00}
\end{align}
respectively, where we define $\mathcal{A}_{\bm{k}} = \bm{v}_{\bm{k}} \bm{A}_0/\mathit{\Omega}$ and replace $\mathit{\Gamma}$ by $\eta$.
Up to the second order in $A_0$, the one-photon Floquet sidebands with the spectral weight $\mathcal{A}_{\bm{k}}^2/4$ appear at $\varepsilon_{\bm{k}} \pm \mathit{\Omega}$.
Using Eqs.~\eqref{eq:expansion_retarded} and \eqref{eq:expansion_lesser}, we find that the time-averaged susceptibility is classified into the following three types: \begin{align}
\chi_{\bm{q}}(\omega)
&= \chi_{\bm{q}}^{\text{base}}(\omega) + \chi_{\bm{q}}^{\text{intra}}(\omega) + \chi_{\bm{q}}^{\text{inter}}(\omega) + \mathcal{O}(A_0^3),
\label{eq:expansion_chi_tot} 
\end{align}
where
\begin{align}
\chi_{\bm{q}}^{\text{base}}(\omega)
&= \frac{2}{N} \sum_{\bm{k}} \Biggl[
\frac{f(\varepsilon_{0,\bm{k}+\bm{q}})-f(\varepsilon_{0,\bm{k}})}{\omega-(\varepsilon_{0,\bm{k}+\bm{q}}-\varepsilon_{0,\bm{k}})+2i\eta}
- \frac{1}{2} \frac{\mathcal{A}_{\bm{k}+\bm{q}}^2 f(\varepsilon_{0,\bm{k}+\bm{q}}) - \mathcal{A}_{\bm{k}}^2 f(\varepsilon_{0,\bm{k}})}{\omega-(\varepsilon_{0,\bm{k}+\bm{q}}-\varepsilon_{0,\bm{k}})+2i\eta}
- \frac{(\mathcal{A}_{\bm{k}+\bm{q}}-\mathcal{A}_{\bm{k}})^2}{2} \frac{f(\varepsilon_{0,\bm{k}+\bm{q}}) - f(\varepsilon_{0,\bm{k}})}{\omega-(\varepsilon_{0,\bm{k}+\bm{q}}-\varepsilon_{0,\bm{k}})+2i\eta}
\Biggr],
\label{eq:expansion_chi_base} \\
\chi_{\bm{q}}^{\text{intra}}(\omega)
&= \frac{2}{N} \sum_{\bm{k}} \Biggl[
\frac{1}{4} \frac{\mathcal{A}_{\bm{k}+\bm{q}}^2 f(\varepsilon_{0,\bm{k}+\bm{q}}+\mathit{\Omega}) - \mathcal{A}_{\bm{k}}^2 f(\varepsilon_{0,\bm{k}}+\mathit{\Omega})}{\omega-(\varepsilon_{0,\bm{k}+\bm{q}}-\varepsilon_{0,\bm{k}})+2i\eta}
+ \frac{1}{4} \frac{\mathcal{A}_{\bm{k}+\bm{q}}^2 f(\varepsilon_{0,\bm{k}+\bm{q}}-\mathit{\Omega}) - \mathcal{A}_{\bm{k}}^2 f(\varepsilon_{0,\bm{k}}-\mathit{\Omega})}{\omega-(\varepsilon_{0,\bm{k}+\bm{q}}-\varepsilon_{0,\bm{k}})+2i\eta}
\Biggr],
\label{eq:expansion_chi_intra} \\
\chi_{\bm{q}}^{\text{inter}}(\omega)
&= \frac{2}{N} \sum_{\bm{k}} \Biggl[
\frac{(\mathcal{A}_{\bm{k}+\bm{q}}-\mathcal{A}_{\bm{k}})^2}{4} \frac{f(\varepsilon_{0,\bm{k}+\bm{q}}) - f(\varepsilon_{0,\bm{k}})}{\omega+\mathit{\Omega}-(\varepsilon_{0,\bm{k}+\bm{q}}-\varepsilon_{0,\bm{k}})+2i\eta}
+ \frac{(\mathcal{A}_{\bm{k}+\bm{q}}-\mathcal{A}_{\bm{k}})^2}{4} \frac{f(\varepsilon_{0,\bm{k}+\bm{q}}) - f(\varepsilon_{0,\bm{k}})}{\omega-\mathit{\Omega}-(\varepsilon_{0,\bm{k}+\bm{q}}-\varepsilon_{0,\bm{k}})+2i\eta}
\Biggr].
\label{eq:expansion_chi_inter}
\end{align}
\end{widetext}
The correction terms proportional to $A_0^2$ are similar to the susceptibility in the high-frequency limit ($\mathit{\Omega}\rightarrow \infty$) in Eq.~\eqref{eq:susceptibility_highfreq} except that the chemical potential $\mu$ or the energy $\omega$ is shifted by $\pm \mathit{\Omega}$.
This implies that the concept of the Fermi-surface nesting is still applicable to the Floquet states, where the energy band in equilibrium ($\varepsilon_{\bm{k}}$) is changed to its time average ($\varepsilon_{0,\bm{k}}$) and some replicas of $\varepsilon_{0,\bm{k}}$ emerge at $\varepsilon_{0,\bm{k}}+n\mathit{\Omega} \ (n=\pm 1,\pm 2,\dots)$.
From the reason which will be given in Sec.~\ref{sec:1d} and illustrated in Fig.~\ref{fig:schema}, three contributions denoted by $\chi^{\text{base}}$, $\chi^{\text{intra}}$, and $\chi^{\text{inter}}$ in Eqs.~\eqref{eq:expansion_chi_base}--\eqref{eq:expansion_chi_inter} are attributed to the electron-hole excitations in the zero-photon Floquet band, those in one of the one-photon Floquet sidebands, and those between the zero-photon band and one-photon Floquet sidebands, respectively.
Note that these expressions are valid for the linearly polarized light with $\bm{A}(t) = \bm{A}_0 \sin \mathit{\Omega} t$, regardless of the lattice structure and the band structure.

\section{Numerical results} \label{sec:results}
In this section, we calculate numerically the static susceptibility in Eq.~\eqref{eq:susceptibility_static} in the two-dimensional square lattice in Sec.~\ref{sec:2d} and the one-dimensional lattice in Sec.~\ref{sec:1d}, and discuss the relation between the susceptibility and the electronic states.
In most of the calculations, we chose $\mathit{\Gamma}=0.05$ and $\beta\rightarrow \infty$, and the sufficiently large $N_p$ for which we have confirmed the convergence.

\subsection{Two-dimensional square lattice} \label{sec:2d}
\begin{figure}[t]\centering
\includegraphics[width=1\columnwidth]{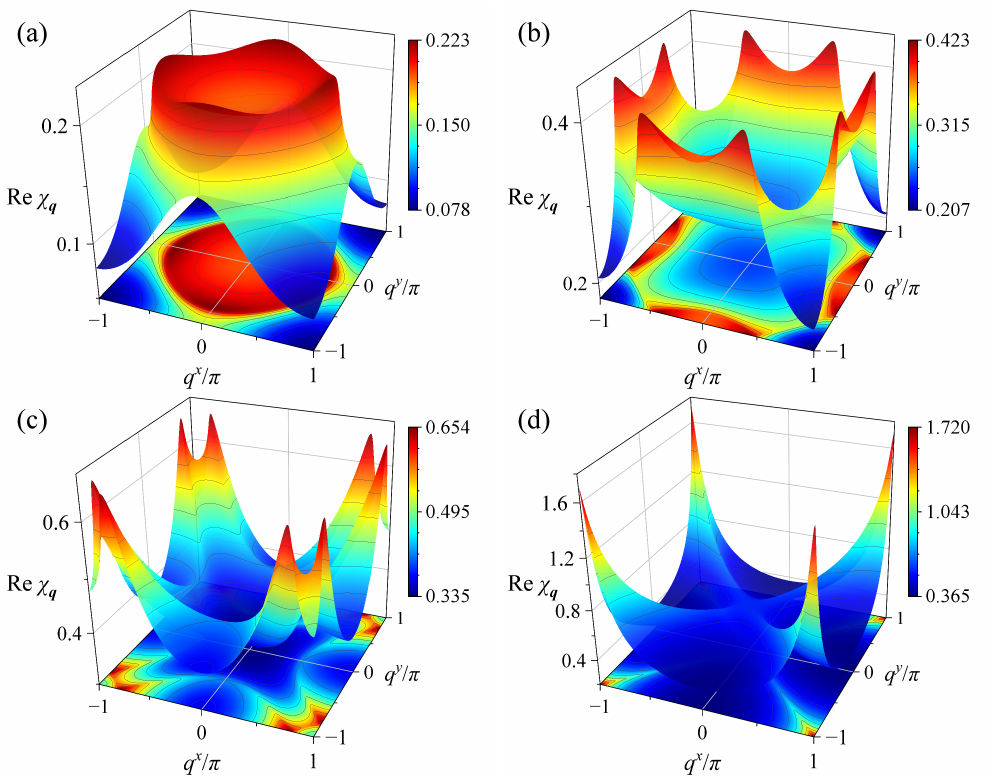}
\caption{The static susceptibility in equilibrium states ($A_0=0$).
The electron density is set to (a)~$n_e=0.25$, (b)~$0.5$, (c)~$0.75$, and (d)~$1$.}
\label{fig:chi_eq_2d}
\end{figure}

We consider the two-dimensional square lattice on which the energy band is defined by
\begin{align}
\varepsilon_{\bm{k}} = -2(\cos k^x + \cos k^y) - \mu,
\end{align}
where $\mu$ is the chemical potential of the system chosen such that the electron density is set to $n_e$.
Energy is measured in units of the absolute value of the nearest-neighbor hopping amplitude.
The number of the lattice sites is $N=256\times 256$.

Figure~\ref{fig:chi_eq_2d} shows the real part of the static susceptibility in the absence of the cw field for different values of $n_e$.
In equilibrium systems, it is widely known that the susceptibility in the momentum space reflects the shape of the Fermi surface, and sharp peaks appear at the nesting vectors $\bm{q}=\bm{Q}$.
When the electron density is small, e.g., $n_e=0.25$ shown in Fig.~\ref{fig:chi_eq_2d}(a), the system is approximately recognized as the free-electron gas with the isotropic Fermi surface, which makes the susceptibility isotropic in the momentum space.
In the case of $n_e=1$, the Fermi surface is perfectly nested with the nesting vector $\bm{Q}=(\pi,\pi)$.
Thus, the sharp peak appears at $\bm{q}=\bm{Q}$ as shown in Fig.~\ref{fig:chi_eq_2d}(d).

\begin{figure}[t]\centering
\includegraphics[scale=0.9]{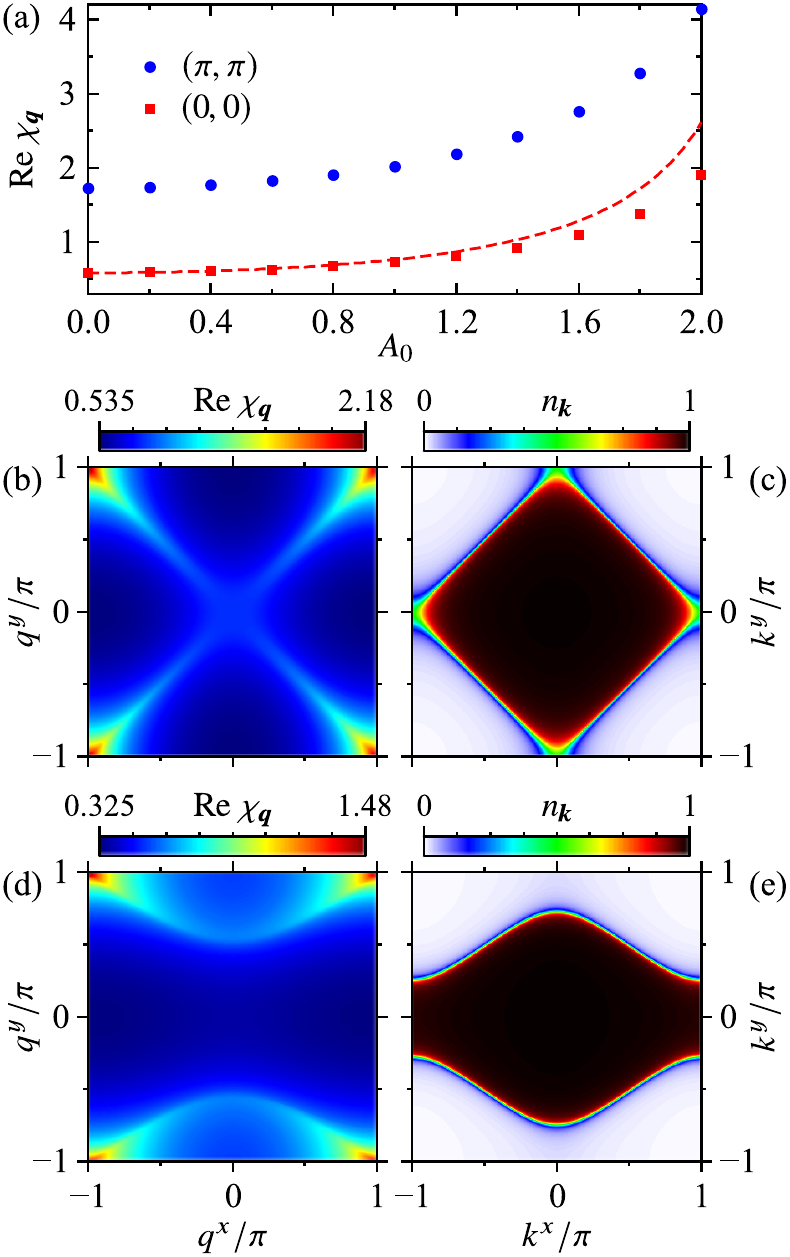}
\caption{(a) The susceptibility at $\bm{q}=(0,0)$ and $(\pi,\pi)$ as a function of $A_0$.
The dashed line represents the density of states in equilibrium divided by $\mathcal{J}_0(A_0)$.
(b)(d) The suscepbility and (c)(e) the momentum distribution in the steady state with $A_0=1.2$ and $\mathit{\Omega}=20$.
The polarization is chosen as (a)--(c) the circular polarization and (d)(e) the linear polarization.
The electron density is $n_e=1$.}
\label{fig:highfreq}
\end{figure}

Now, we show the susceptibility in the presence of the cw field.
First, we focus on the susceptibility in a high-frequency regime where $\mathit{\Omega}$ is larger than the bandwidth ($=8$).
The susceptibility is written as Eq.~\eqref{eq:susceptibility_highfreq}, where $\varepsilon_{0,\bm{k}}$ is the time average of $\varepsilon_{\bm{k}-\bm{A}(t)}$ during the period $T$.
When the circularly polarized light is applied, $\varepsilon_{0,\bm{k}}$ is given by
\begin{align}
\varepsilon_{0,\bm{k}} = -2 \mathcal{J}_0(A_0)(\cos k^x + \cos k^y) - \mu,
\label{eq:band_circ}
\end{align}
where $\mathcal{J}_n$ is the $n$th-order Bessel function of the first kind, indicating the DL effect, i.e., a reduction in the electron bandwidth~\cite{Dunlap1986,Holthaus1992,Grossmann1991,Kayanuma2008,Kawakami2018}.
We note that the circularly polarized light acts simply as an isotropic external field in the present free-electron model where the spin-orbit coupling and the Zeeman term are not taken into account.
When the circularly polarized light is applied, the susceptibility shown in Fig.~\ref{fig:highfreq}(b) is qualitatively the same as the results in the equilibrium state presented in Fig.~\ref{fig:chi_eq_2d}(d).
An increase in $\chi_{\bm{q}}$ is found in the whole $\bm{q}$ region.
Figure~\ref{fig:highfreq}(a) shows the susceptibility at $\bm{q}=(0,0)$ and $(\pi,\pi)$ as a function of the amplitude.
It is found that the susceptibility increases monotonically with increasing $A_0$.
The dashed curve in Fig.~\ref{fig:highfreq}(a) shows the density of states in equilibrium ($A_0=0$) divided by the Bessel function, $\left. n(0) \right\vert_{A_0=0}/\mathcal{J}_0(A_0)$, which fits the data of $\bm{q}=(0,0)$ quite well, altough some deviations are seen for $A_0=1.6\text{--}2$ due to the finite $\mathit{\Gamma}$.
Thus the increase in $\chi_{\bm{q}}$ in the high-frequency regime is ascribed to the DL effect.

In addition to the DL effect discussed above, the shape of the Fermi surface and thus the nesting vector are controlable by applying the linearly polarized light given by Eq.~\eqref{eq:def_linear_pol}.
The time-averaged energy band is given by
\begin{align}
\varepsilon_{0,\bm{k}}
= -2[\mathcal{J}_0(A_0^x)\cos k^x + \mathcal{J}_0(A_0^y)\cos k^y] - \mu.
\label{eq:band_linear}
\end{align}
We consider the case of $(A_0^x,A_0^y)=(1.2,0)$ as an example.
The momentum distribution function in Fig.~\ref{fig:highfreq}(e) shows a remarkable difference from that in the circularly polarized light (Fig.~\ref{fig:highfreq}(c)).
Modification of the Fermi surface brings about the anisotropic susceptibility with the two-fold symmetry shown in Fig.~\ref{fig:highfreq}(d).

\begin{figure}[t]\centering
\includegraphics[scale=0.9]{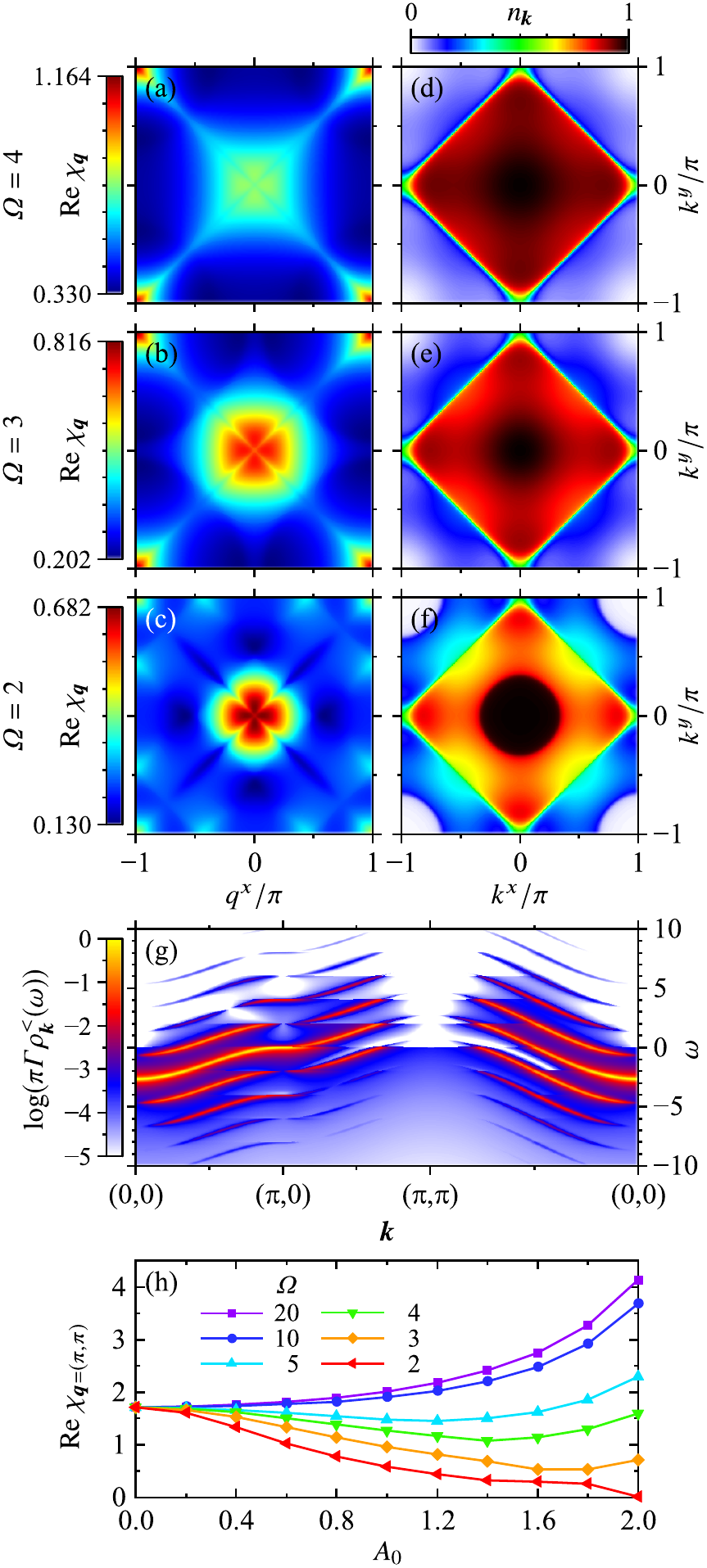}
\caption{(a)--(c) The susceptibility and (d)--(f) the momentum distribution for $A_0=1.2$ and $\mathit{\Omega}=4,3,2$ (top to bottom).
(g) The spectral function $\rho_{\bm{k}}^{<}(\omega)$ for $A_0=1.2$ and $\mathit{\Omega}=2$.
(h) The susceptibility at $\bm{q}=(\pi,\pi)$ as a function of $A_0$ for different values of $\mathit{\Omega}$.
The circularly polarized light is applied.
The electron density is $n_e=1$.}
\label{fig:freqdep}
\end{figure}

\begin{figure*}[t]\centering
\includegraphics[width=1\hsize]{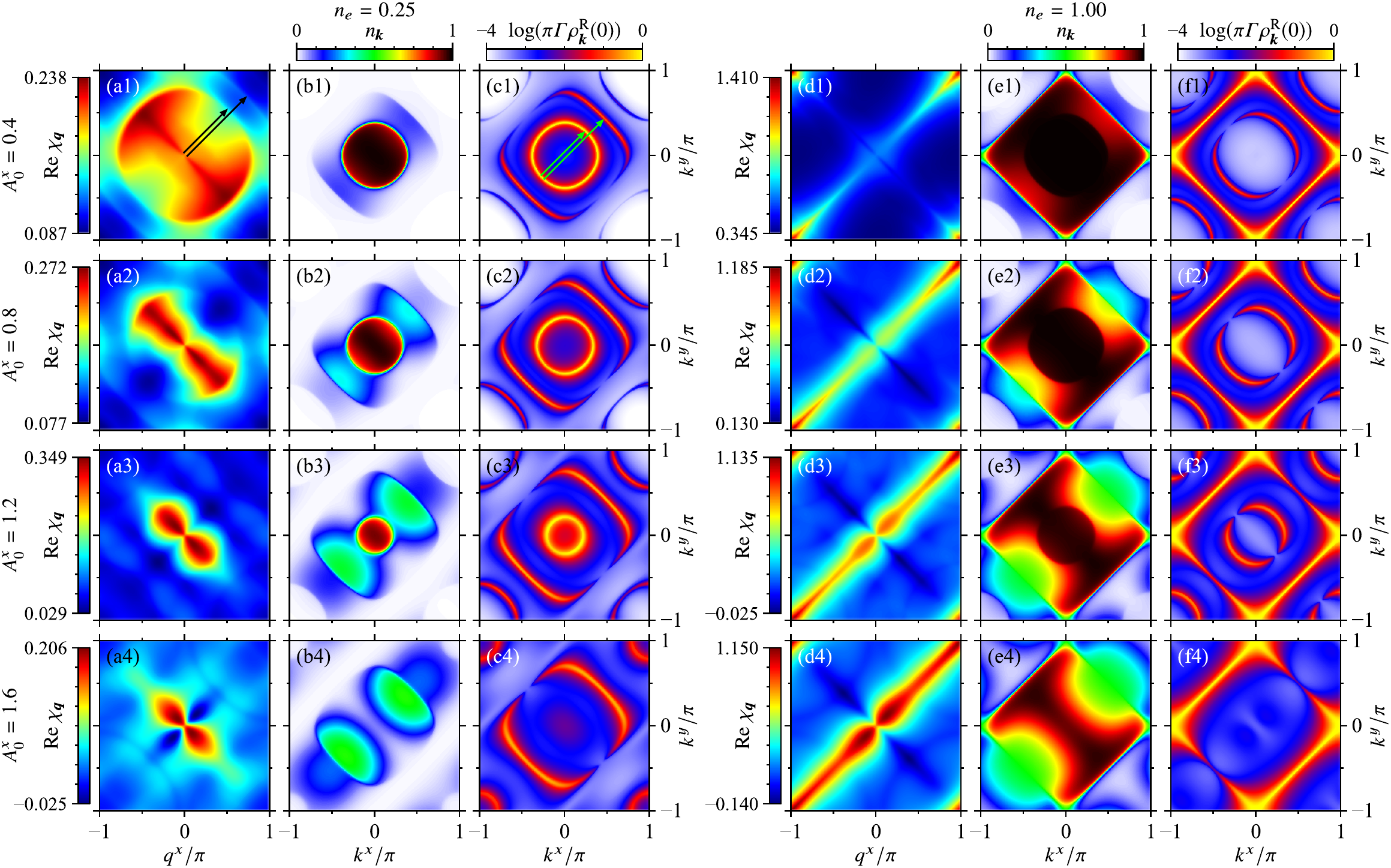}
\caption{(a)(d) The susceptibility, (b)(e) the electron distribution function, and (c)(f) the spectral function at $\omega=0$, under the linearly polarized light with $A_0^x=A_0^y=0.4,0.8,1.2,1.6$ (top to bottom).
Arrows in (a1) and (c1) represent $\bm{q}/\pi=(0.5,0.5)$ and $\bm{q}/\pi=(0.7.0.7)$ as guides for the eye (see text).
The cw frequency is $\mathit{\Omega}=2$ and the electron density is (a)--(c) $n_e=0.25$ and (d)--(f) $n_e=1$.}
\label{fig:lowfreq}
\end{figure*}

Next, we present the cw-field frequency dependence of the susceptibility.
Figures~\ref{fig:freqdep}(a)--\ref{fig:freqdep}(f) show the susceptibility and the momentum distribution function in the steady state under the circularly polarized light for different values of $\mathit{\Omega}$.
The chemical potential is set to $\mu=0$, which keeps the system half-filled ($n_e=1$) for any $A_0$ and $\mathit{\Omega}$ owing to the electron-hole symmetry.
As for $\mathit{\Omega}=3$ and $4$ (Figs.~\ref{fig:freqdep}(a) and \ref{fig:freqdep}(b)), $\chi_{\bm{q}}$ and $n_{\bm{k}}$ are qualitatively the same as those at $\mathit{\Omega}=20$ shown in Figs.~\ref{fig:highfreq}(b) and \ref{fig:highfreq}(c), although $n_{\bm{k}}$ slightly spreads outside the Fermi surface (given by $|k^x|+|k^y|=\pi$) and $\chi_{\bm{q}}$ increases around $\bm{q}=(0,0)$.
In this sense, at $\mathit{\Omega}=3\text{--}4$ and $A_0^x=1.2$, the system is recognized to be in the high-frequency regime even though $\mathit{\Omega}$ is smaller than the bandwidth.
In contrast, at $\mathit{\Omega}=2$ (Fig.~\ref{fig:freqdep}(c)), the peak intensity of $\chi_{\bm{q}}$ at $\bm{q}=(0,0)$ is higher than that at $\bm{q}=(\pi,\pi)$.
Emergent peaks are found in $\chi_{\bm{q}}$ at $\bm{q}/\pi=(\pm 0.5, 1)$ and $(1,\pm 0.5)$, and two other discontinous lines appear on circles centered at $\bm{q}=(0,0)$ and $(\pi,\pi)$ with radius $0.33\pi$ in addition to the one that arises from the zero-photon Fermi surface.
In Fig.~\ref{fig:freqdep}(g), the spectral function $\rho_{\bm{k}}^{<}(\omega)$ shows that not only the zero-photon band but also one-photon bands cross the zero energy, which induces the discontinuity in $n_{\bm{k}}$.
The amplitude dependence of the susceptibility at $\bm{q}=(\pi,\pi)$ is plotted in Fig.~\ref{fig:freqdep}(h).
The susceptibility increases monotonically with increasing $A_0$ for $\mathit{\Omega}=10$ and $20$, whereas it exhibits non-monotonic dependence on $A_0$ for $\mathit{\Omega} = 2\text{--}5$.
Since DL tends to enhance the susceptibility, the reduction in $\chi_{\bm{q}}$ is ascribed to the nonthermal electron distribution function.
In particular, when the Floquet sidebands with energy $\varepsilon_{0,\bm{k}}-l\mathit{\Omega}$ have the ``Fermi surfaces'', $n_{\bm{k}}$ and $\chi_{\bm{q}}$ undergo the qualitative change from those in the high-frequency regime.

Let us consider the low-frequency regime ($\mathit{\Omega}=2$), where the some Floquet bands cross the Fermi level of the bath.
Figure~\ref{fig:lowfreq} shows the amplitude dependence of the susceptibility, the electron distribution function, and the spectral function at $\omega=0$.
The electron density is chosen to $n_e=0.25$ and $1$, and the light is linearly polarized as $A_0^x=A_0^y$.
In the case of $n_e=0.25$ (Figs.~\ref{fig:lowfreq}(a1)--\ref{fig:lowfreq}(c4)), not only the Fermi surface of the zero-photon band but also the ones of one-photon and two-photon bands appear in $\rho_{\bm{k}}^{\mathrm{R}}(0)$, which gives rise to the discontinuity of $n_{\bm{k}}$ at the Fermi surfaces.
The zero-photon Fermi surface is almost isotropic, and shrinks with increasing $A_0$ due to the appearance of the Floquet sidebands in addition to DL.
The electron distribution $n_{\bm{k}}$ spreads outside the Fermi surface along the light-polarization direction, whereas it does not along the direction perpendicular to the light.
From the results in Sec.~\ref{sec:expansion}, the suscepbitility in the steady state is approximately understood from Eq.~\eqref{eq:susceptibility_lowfreq}.
Here, we rewrite Eq.~\eqref{eq:susceptibility_lowfreq} as
\begin{align}
\chi_{\bm{q}}(\omega)
= \frac{2}{N} \sum_{\bm{k}} \frac{n_{\bm{k}+\bm{q}}-n_{\bm{k}}}{\omega-(\varepsilon_{\bm{k}+\bm{q}}-\varepsilon_{\bm{k}})+2i\eta},
\label{eq:susceptibility_eq}
\end{align}
where the Fermi--Dirac function $f(\varepsilon_{\bm{k}})$ is replaced by the nonequilibrium electron distribution function $n_{\bm{k}}$.
Equation~\eqref{eq:susceptibility_eq} implies that characteristic structure of $\chi_{\bm{q}}$ appears at a ``nesting vector'' $\bm{q}=\bm{Q}$ that connects the two points of the Fermi surfaces of the Floquet bands.
At $\bm{q}=\bm{Q}$, $\varepsilon_{\bm{k}+\bm{Q}}-\varepsilon_{\bm{k}}$ in the denominator is regarded as zero and $\vert n_{\bm{k}+\bm{Q}}-n_{\bm{k}} \vert$ in the numerator is large.
In equilibrium, a region in the momentum space in which $\chi_{\bm{q}}$ shows a large value
is nearly a circle with radius $2k_{\mathrm{F}}$, where $k_{\mathrm{F}} = 0.4\pi$ is the Fermi wavenumber in the case of $n_e=0.25$, as shown in Fig.~\ref{fig:chi_eq_2d}(a).
When the amplitude is small ($A_0^x=A_0^y = 0.4$ and $0.8$), the circle shrinks and the intensity around $\bm{q}/\pi=(0.5,0.5)$ ($\bm{q}/\pi=(0.7,0.7)$) decreases (increases) compared to the case of $A_0=0$, reflecting the changes in $n_{\bm{k}}$ and $\rho_{\bm{k}}^{\mathrm{R}}(\omega)$.
In the case of $A_0^x=A_0^y > 0.8$, where the system is far beyond the second-order regime described by Eqs.~\eqref{eq:expansion_chi_tot}--\eqref{eq:expansion_chi_inter}, two maxima and two minima of $\chi_{\bm{q}}$ approach $\bm{q}=(0,0)$, while the correspondence between $n_{\bm{k}}$ and $\chi_{\bm{q}}$ is difficult to find in Figs.~\ref{fig:lowfreq}(a3) and \ref{fig:lowfreq}(a4).

The susceptibility, the momentum distribution function, and the Fermi surface in the half-filled ($n_e=1$) system are shown in Figs.~\ref{fig:lowfreq}(d1)--\ref{fig:lowfreq}(f4).
A major difference from the case of $n_e=0.25$ is the presence of the zero-photon band and the nesting vector $\bm{Q}=(-\pi,\pi)$ for large $A_0$ due to the electron-hole symmetry.
The linearly-polarized light spreads $n_{\bm{k}}$ along the polarization direction, which reduces the spectral weight at $\omega=0$, whereas $n_{\bm{k}}$ on the perpendicular direction is not affected by the light.
Accordingly, the susceptibility exhibits the one-dimensional-like structure with increasing $A_0$.

\begin{figure}[t]\centering
\includegraphics[scale=0.9]{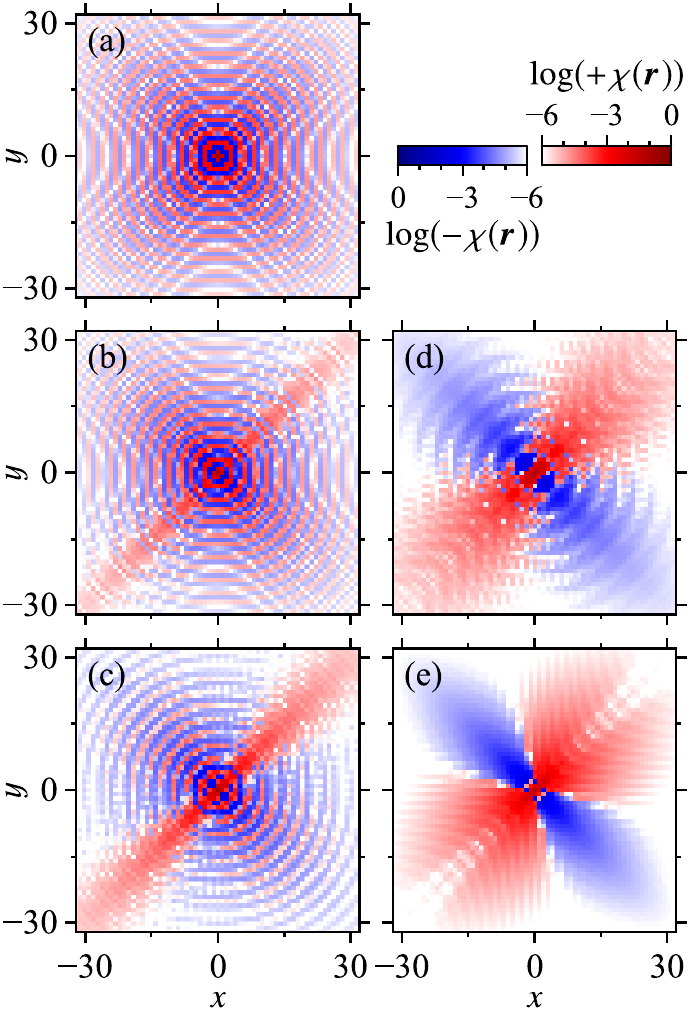}
\caption{The susceptibility in the real space, $\chi(\bm{r}_i-\bm{r}_j) \equiv (\chi_{ij})_{00}(0)$, for (a)~$A_0^x=A_0^y=0$, (b)~$0.4$, (c)~$0.8$, (d)~$1.2$, and (e)~$1.6$.
The parameter values are chosen to $\mathit{\Omega}=2$ and $n_e=0.25$.}
\label{fig:realspace}
\end{figure}

The Fourier transformation of $\chi_{\bm{q}}$ with respect to $\bm{q}$ given by
\begin{align}
\chi(\bm{r}) = \frac{1}{N} \sum_{\bm{q}} e^{i\bm{q}\bm{r}} \chi_{\bm{q}}
\label{eq:susceptibility_realspace2}
\end{align}
describes the interaction between the two magnetic impurities separated by a distance $\bm{r}$ as mentioned in Sec.~\ref{sec:susceptibility}.
Figure~\ref{fig:realspace} shows $\chi(\bm{r})$ for $A_0^x=A_0^y=0\text{--}1.6$ in the case of $n_e=0.25$ and $\mathit{\Omega}=2$.
In equilibrium, $\chi(\bm{r})$ presented in Fig.~\ref{fig:realspace}(a) is four-fold symmetric and shows oscillating behavior with a period of $2\pi/(2k_{\mathrm{F}}) \approx 2$ (sites).
As $A_0$ increases, $\chi(\bm{r})$ is modulated in accordance with the change in $\chi_{\bm{q}}$ with the two-fold symmetry.
In particular, a FM correlation along the light polarization is enhanced.
For large $A_0$, the oscillation is no longer observed and the correlation exhibits short-range behavior.
In a long-range region for $r=\Vert\bm{r}\Vert \gtrsim \ell$, where $\ell =v_{\mathrm{F}}/\mathit{\Gamma}$ is the mean free path of the electrons with the Fermi velocity $v_{\mathrm{F}}$, the correlation $\chi(\bm{r})$ decays exponentially with respect to $\bm{r}$ (not shown).

\begin{figure}[t]\centering
\includegraphics[width=1\columnwidth]{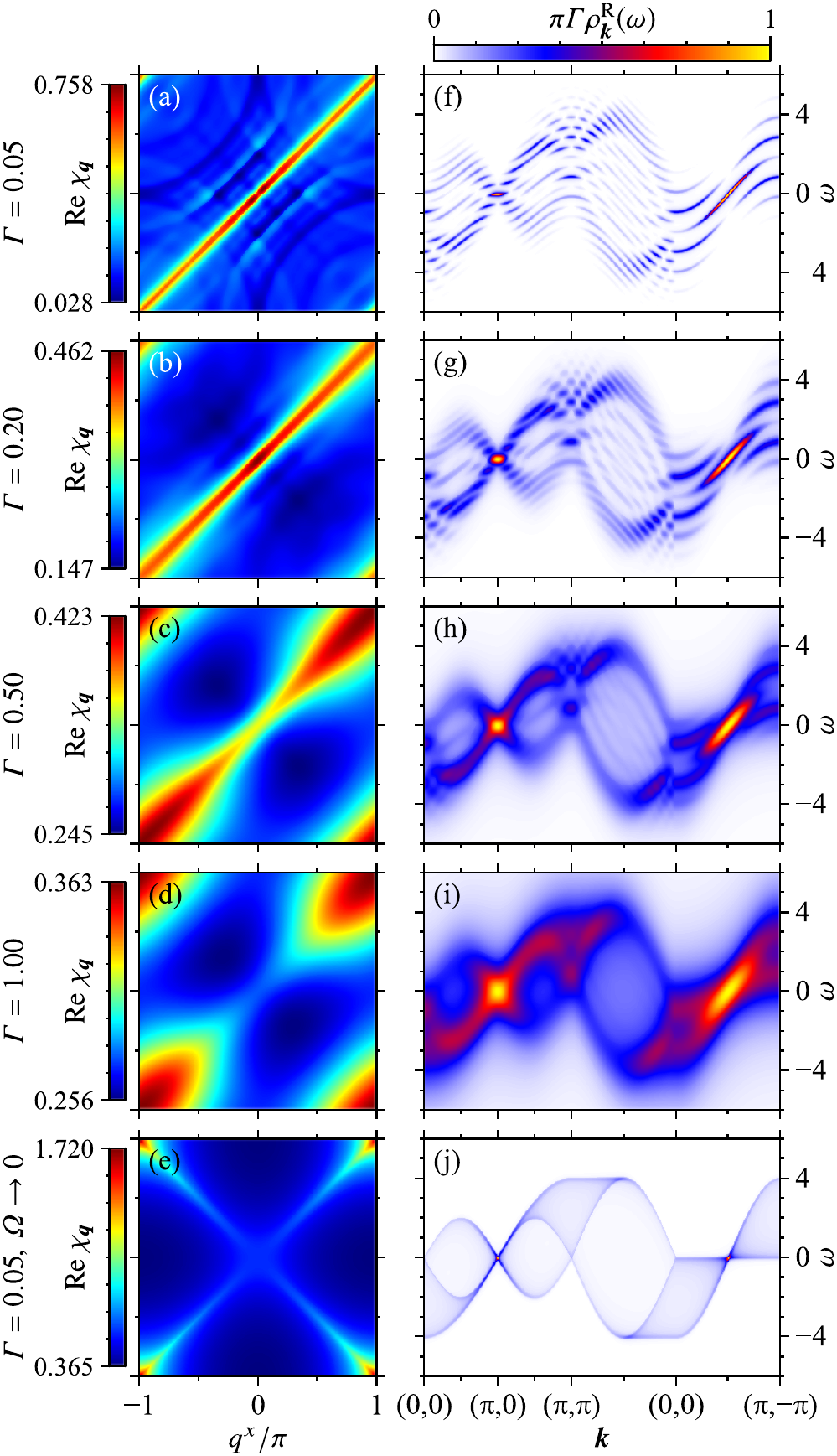}
\caption{(a)--(e) The susceptibility and (f)--(j) the normalized spectral function under the linearly polarized light with $A_0^x=A_0^y=\pi/2$.
The coupling strength between the system and the bath is $\mathit{\Gamma}=0.05,0.2,0.5,1,0.05$ (top to bottom).
The frequency of light is chosen to $\mathit{\Omega}=0.5$ except for (e) and (j).
Result in (e) is the same with Fig.~\ref{fig:chi_eq_2d}(d), and the spectral function in (j) is obtained from Eq.~\eqref{eq:green_retarded_lowfreq}, where $\mathit{\Omega}\rightarrow 0$.
The dimension of the Floquet space is set to $N_p=4096$ in (j) and $N_p=40$ in the others.}
\label{fig:etadep}
\end{figure}

Finally, we consider the low-frequency limit ($\mathit{\Omega}\rightarrow 0$) and discuss the relation between $\mathit{\Omega}$ and $\mathit{\Gamma}$.
As discussed in Secs.~\ref{sec:model} and \ref{sec:susceptibility}, in the limit of $\mathit{\Omega}\rightarrow 0$, the Green functions are given by Eqs.~\eqref{eq:green_retarded_lowfreq}--\eqref{eq:green_lesser_lowfreq_continuous};
thus, the susceptibility is reduced to the equilibrium one in Eq.~\eqref{eq:susceptibility_lowfreq}.
We examine a crossover of the susceptibility from $\mathit{\Omega}>\mathit{\Gamma}$ to $\mathit{\Omega} \lesssim \mathit{\Gamma}$, varying $\mathit{\Gamma}$ instead of $\mathit{\Omega}$.
Figures~\ref{fig:etadep}(a)--\ref{fig:etadep}(d) and Figs.~\ref{fig:etadep}(f)--\ref{fig:etadep}(i) present the susceptibility and the spectral function, respectively, for $\mathit{\Gamma}=0.05\text{--}1$ and $\mathit{\Omega}=0.5$.
Figure~\ref{fig:etadep}(j) shows the spectral function in the case of $\mathit{\Omega}\rightarrow 0$ calculated from Eq.~\eqref{eq:green_retarded_lowfreq}, and Fig.~\ref{fig:etadep}(e) displays the corresponding susceptibility that is already shown in Fig.~\ref{fig:chi_eq_2d}(d), for comparison.
At $\mathit{\Gamma}=0.05 \ (\ll \mathit{\Omega})$, the Floquet bands are distinguishable with each other.
As $\mathit{\Gamma}$ increases, these Floquet bands merge together and the continuum is formed, where the spectral intensity is qualitatively the same as that of the low-frequency limit ($\mathit{\Omega}\rightarrow 0$) shown in Fig.~\ref{fig:etadep}(j).
The peak in the susceptibility is broadened on a line connecting $\bm{q}=(\pi,\pi)$ and $(-\pi,-\pi)$, and is diminished around $\bm{q}=(0,0)$.
Therefore, the susceptibility has the peak at $\bm{q}=(\pi,\pi)$ similarly to the equilibrium susceptibility when $\mathit{\Omega}$ is much smaller than $\mathit{\Gamma}$.

\subsection{One-dimensional lattice} \label{sec:1d}

\begin{figure}[t]\centering
\includegraphics[scale=1]{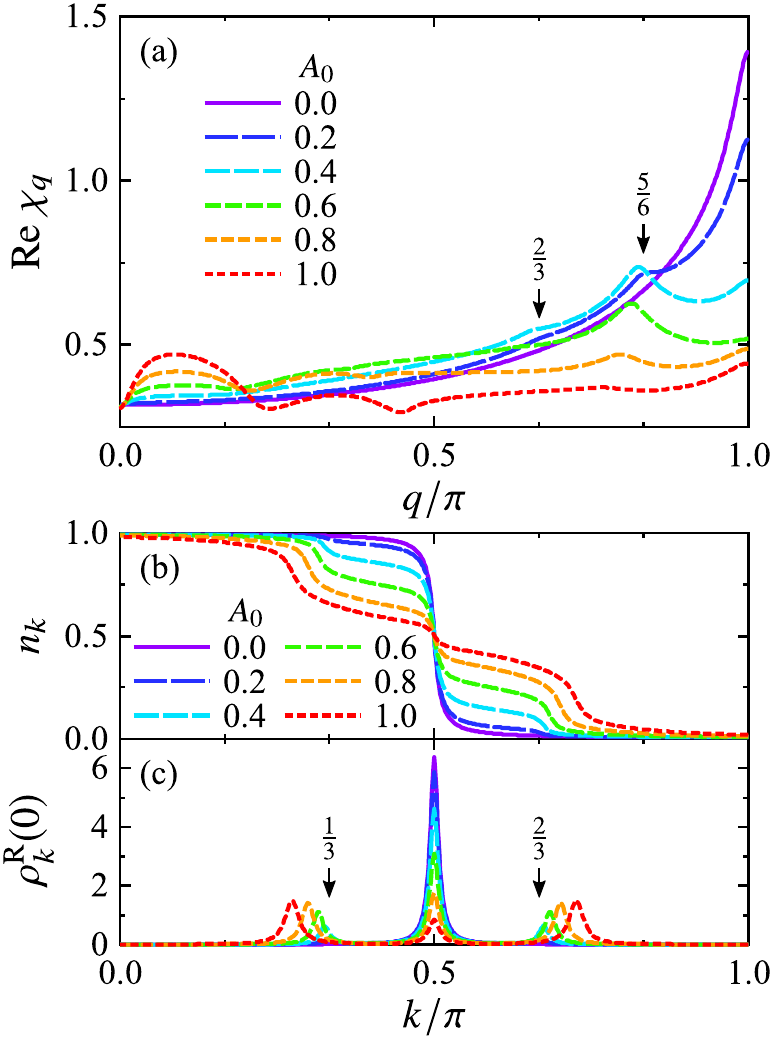}
\caption{(a) The susceptibility, (b) the momentum distribution function, and (c) the spectral function at $\omega=0$ in the one-dimensional system for $A_0=0\text{--}1$.
The cw frequency and the electron density are set to $\mathit{\Omega}=1$ and $n_e=1$, respectively.}
\label{fig:1d_chi}
\end{figure}

In this section, we consider the one-dimensional lattice with the energy band given by
\begin{align}
\varepsilon_{k} = -2\cos k -\mu,
\label{eq:band_1d}
\end{align}
where the nearest-neighbor hopping amplitude is taken to be unity.
We focus on the half-filled case with $\mu=0$ and $n_e=1$.
The number of the lattice sites is $N=1024$ and the light frequency is set to $\mathit{\Omega}=1$ (i.e., the low-frequency regime).
In Fig.~\ref{fig:1d_chi}, the susceptibility, the momentum distribution function, and the spectral function are plotted for different values of $A_0$.
The Fermi wavenumber $k_{\mathrm{F}}$ is $\pi/2$ in the case of $A_0=0$, where a peak in $\rho_{k}^{\mathrm{R}}(0)$ and a jump in the $n_{\bm{k}}$ are seen.
Correspondingly, the susceptibility has a peak at the nesting vector $Q=2k_{\mathrm{F}}=\pi$ as shown in Fig.~\ref{fig:1d_chi}(a).
In the presence of the cw field, additional peaks associated with the one-photon bands appear at $k=\cos^{-1}(\pm 1/2)=\pi/2\mp \pi/6$, when $A_0$ is small.
These peaks are separated from the peak at $k=\pi/2$ with increasing $A_0$ due to the DL effect.
The susceptibility at $q=\pi$ decreases, and that at $q=5\pi/6$ and $2\pi/3$ increase for $A_0\lesssim 0.4$.

\begin{figure}[t]\centering
\includegraphics[scale=1]{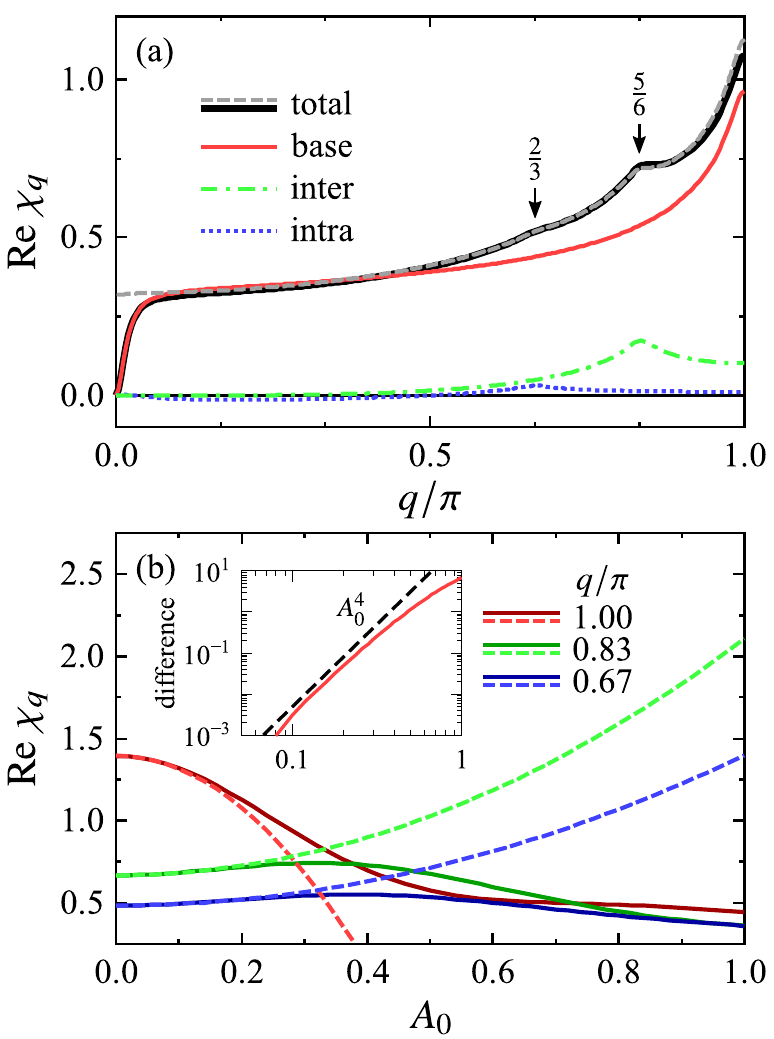}
\caption{(a) The suscepbility obtained from Eq.~\eqref{eq:susceptibility_floquet} (dashed line) and that from Eq.~\eqref{eq:expansion_chi_tot} (bold line).
The three contributions given by Eqs.~\eqref{eq:expansion_chi_base}--\eqref{eq:expansion_chi_inter} are also plotted by thin line, dotted line, and dot-dash line, respectively.
(b) The susceptibility at $q/\pi=1, 0.83, 0.67$ as a function of $A_0$, obtained from Eq.~\eqref{eq:susceptibility_floquet} (solid lines) and that from Eq.~\eqref{eq:expansion_chi_tot} (dashed lines).
Inset shows difference between the numerically exact susceptibility in Eq.~\eqref{eq:susceptibility_floquet} and the approximated one in Eq.~\eqref{eq:expansion_chi_tot}, at $q=\pi$ as solid line.
Dashed line in inset represents a slope of $A_0^4$ as a guide for the eye.
The frequency of light and the electron density is set to $\mathit{\Omega}=1$ and $n_e=1$, respectively.}
\label{fig:1d_expansion}
\end{figure}

The wavenumbers $q=5\pi/6$ and $q=2\pi/3$, at which the additional peaks appear in $\chi_q$, turn out to be ``inter-Floquet-band'' and ``intra-Floquet-band'' nesting vectors, respectively, from the series expansion of $\chi_{\bm{q}}$ in Eqs.~\eqref{eq:expansion_chi_tot}--\eqref{eq:expansion_chi_inter}.
Figure~\ref{fig:1d_expansion}(a) shows the numerical results
of the susceptibility calculated exactly in the numerical sense from Eq.~\eqref{eq:susceptibility_floquet} (dashed line), and the susceptibility calculated from Eqs.~\eqref{eq:expansion_chi_tot}--\eqref{eq:expansion_chi_inter} (bold line).
These are in good agreement with each other~\footnote{There is a deviation between the numerically exact susceptibility and the approximated one at $q=0$.
This is because the former is computed via the fast Fourier transformation of Eq.~\eqref{eq:susceptibility_floquet}, whereas the latter is directly evaluated from Eqs.~\eqref{eq:expansion_chi_base}--\eqref{eq:expansion_chi_inter}, which vanish since all of the numerators are zero for $q=0$.}.
It is found that the shoulders at $q=5\pi/6$ and $2\pi/3$ are ascribed to the ``inter-Floquet-band'' contribution given in Eq.~\eqref{eq:expansion_chi_inter} (dot-dash line) and the ``intra-Floquet-band'' contribution given in Eq.~\eqref{eq:expansion_chi_intra} (dotted line), respectively.
In Fig.~\ref{fig:1d_expansion}(b), we plot $\chi_{q}$ at three characteristic wavenumbers, $q/\pi = 1, 0.83, 0.67$, as a function of $A_0$ (dashed lines), which coincide with the numerically exact susceptibilities (solid lines) for small $A_0$.
The inset of Fig.~\ref{fig:1d_expansion}(b) shows that the difference of $\chi_{q=\pi}^{\text{base}}$ from $\chi_{q=\pi}$ is proportional to $A_0^4$, as expected.

\begin{figure}[t]\centering
\includegraphics[width=1\columnwidth]{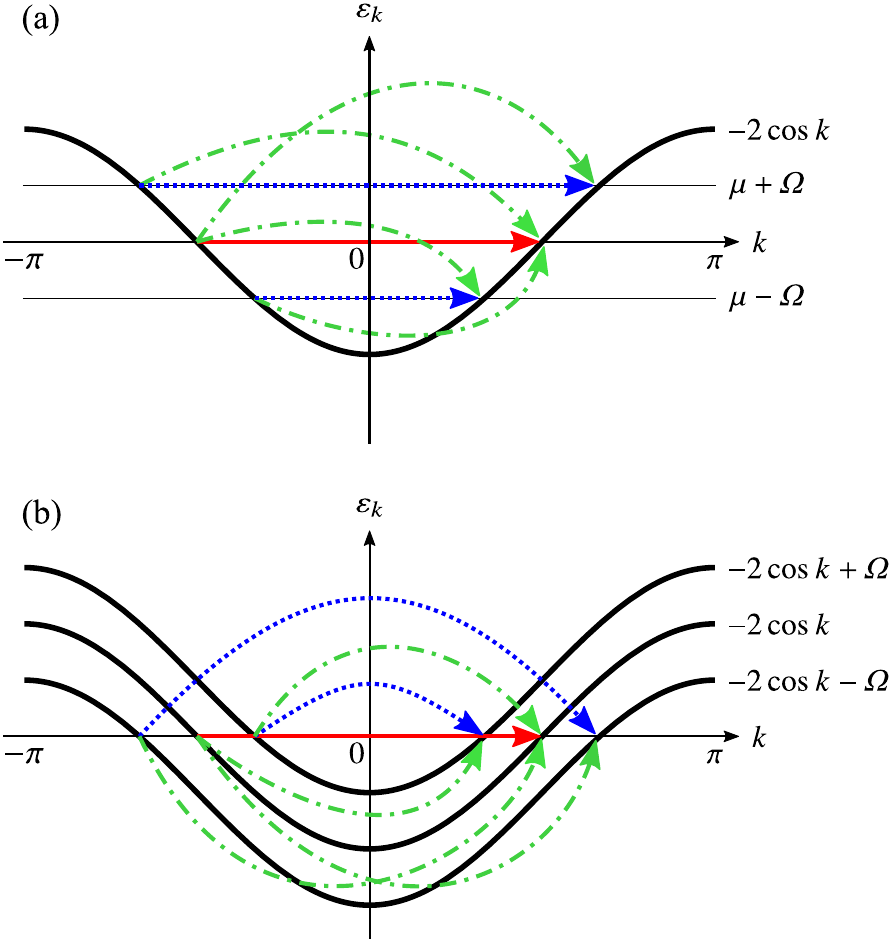}
\caption{(a) The schematic diagram of the energy band and the electron-hole excitations associated with $\chi_q^{\text{base}}$ (solid arrow), $\chi_q^{\text{intra}}$ (dotted arrows), and $\chi_q^{\text{inter}}$ (dot-dash arrows).
(b) An interpretation of the three types of the contribution based on the Floquet-band picture, where the Floquet sidebands appear around the original zero-photon band with spacing $\mathit{\Omega}$.
The dotted arrows represent the electron-hole excitations in one of the Floquet sidebands (the intra-Floquet-band excitations), and the dot-dash arrows indicate the electron-hole excitations between the zero-photon band and one of the one-photon bands (the inter-Floquet-band excitations).}
\label{fig:schema}
\end{figure}

In order to clarify the origin of the emergent peaks (shoulders) in the susceptibility, we derive more simplified expressions of Eqs.~\eqref{eq:expansion_chi_base}--\eqref{eq:expansion_chi_inter}.
We consider the one-dimensional system and assume $\mathcal{A}_{k} = \pm v_{\mathrm{F}}A_0/\mathit{\Omega} = \mp \mathcal{A}_{k+q}$ for large $q$.
This leads to the following expressions:
\begin{align}
\chi_{q}^{\text{base}}(\omega)
&= {\left( 1-\frac{5}{2} \mathcal{A}_{\mathrm{F}}^2 \right)} \chi_{q}^{(0)}(\omega),
\label{eq:expansion_chi_base_1d} \\
\chi_{q}^{\text{intra}}(\omega)
&= \frac{\mathcal{A}_{\mathrm{F}}^2}{4} {\left[
\left.\chi_{q}^{(0)}(\omega)\right\vert_{\mu\rightarrow \mu-\mathit{\Omega}}
+ \left.\chi_{q}^{(0)}(\omega)\right\vert_{\mu\rightarrow \mu+\mathit{\Omega}}
\right]},
\label{eq:expansion_chi_intra_1d} \\
\chi_{q}^{\text{inter}}(\omega)
&= \mathcal{A}_{\mathrm{F}}^2 {\left[
\chi_{q}^{(0)}(\omega+\mathit{\Omega}) + \chi_{q}^{(0)}(\omega-\mathit{\Omega})
\right]},
\label{eq:expansion_chi_inter_1d}
\end{align}
with $\mathcal{A}_{\mathrm{F}} = v_{\mathrm{F}}A_0/\mathit{\Omega}$.
Here, $\chi_{q}^{(0)}(\omega)$ is the susceptibility in the high-frequency limit ($\mathit{\Omega}\rightarrow\infty$) defined by Eq.~\eqref{eq:susceptibility_highfreq}.
The coefficient of $\chi_q^{\text{inter}}$ is four times larger than $\chi_q^{\text{intra}}$.
Note that $\chi_{q}^{\text{inter}}$ vanishes for the small $\bm{q}$ such that $\mathcal{A}_{\bm{k}}\approx \mathcal{A}_{\bm{k}+\bm{q}}$, because of the factor $(\mathcal{A}_{\bm{k}+\bm{q}}-\mathcal{A}_{\bm{k}})^2$ in Eq.~\eqref{eq:expansion_chi_inter}.
The schematic illustration of the above equations is shown in Fig.~\ref{fig:schema}.
Equation~\eqref{eq:expansion_chi_intra_1d} consists of the two kinds of $\chi_q^{(0)}$ whose chemical potentials are shifted by $\pm\mathit{\Omega}$.
These describe the nonequilibrium susceptibility that originates from the intra-Floquet-band electron-hole excitations.
On the other hand, Eq.~\eqref{eq:expansion_chi_inter_1d} is composed of $\chi_q^{(0)}(\omega+\mathit{\Omega})$ and $\chi_q^{(0)}(\omega-\mathit{\Omega})$, which are regarded as the inter-Floquet-band electron-hole excitations.

\begin{figure}[t]\centering
\includegraphics[scale=1]{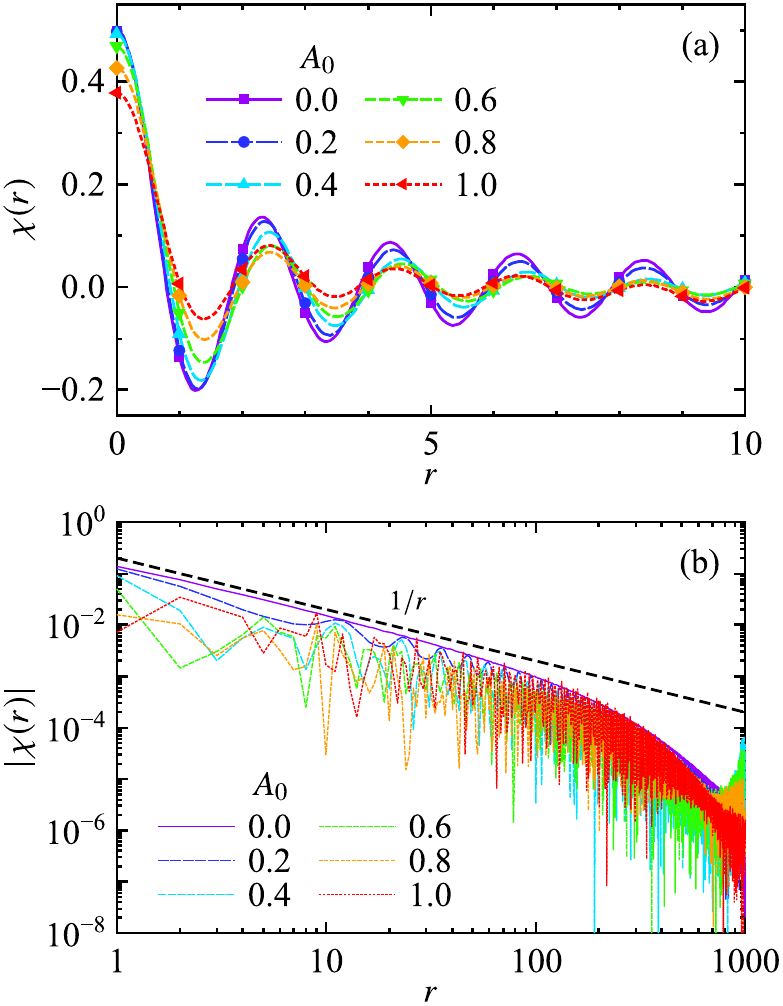}
\caption{The susceptibility $\chi(r)$ for (a) $r\in[0,10]$ and (b) $r\in [1,10^3]$.
The symbols in (a) indicate $\chi(r)$ at integer numbers of $r$.
The bold curves in (a) and the bold dashed line in (b) are guides for the eye.
The cw-field frequency is set to $\mathit{\Omega}=1$.}
\label{fig:1d_realspace}
\end{figure}

Finally, we discuss the susceptibility in the real space.
We calculate $\chi_q$ in Eq.~\eqref{eq:susceptibility_floquet} and perform the Fourier transformation in Eq.~\eqref{eq:susceptibility_realspace2}.
In order to discuss the long-range behavior of $\chi(r)$, the system-bath coupling $\mathit{\Gamma}$ is set to $\mathit{\Gamma}=0.005$, which governs the mean free path as $\ell = v_{\mathrm{F}}/\mathit{\Gamma} = 400$, and the system size is $N=2048 \ (\gg \ell)$.
Figure~\ref{fig:1d_realspace} shows $\chi(r)$ both on a linear scale and on a logarithmic scale.
In the equilibrium system ($A_0=0$), $\chi(q)$ oscillates with a period of $2\pi/(2k_{\mathrm{F}}) = 2$ and decays as $\chi(r) \sim 1/r$, reflecting the one-dimensional nature.
In the presence of the cw field, the period of the oscillation is slightly modulated in accordance with the appearance of the intra- and inter-Floquet-band peaks in $\chi_{q}$.
However, the power-law behavior at long distances does not change since the Fermi-surface nesting remains in the case of the one-dimensional lattice.

\section{Summary} \label{sec:summary}
In this paper, we have studied the spin and charge susceptibilities in the photoinduced Floquet states.
Considering the electon system coupled to the fermionic bath under linearly or circularly polarized light, we derived the formula of the susceptibility in Eq.~\eqref{eq:susceptibility_floquet}.
In the high-frequency limit ($\mathit{\Omega}\rightarrow \infty$), the susceptibility is written as Eq.~\eqref{eq:susceptibility_highfreq}, which is similar to the equilibrium susceptibility except that the energy band $\varepsilon_{\bm{k}}$ is replaced by the time-averaged energy band $\varepsilon_{0,\bm{k}}$.
Due to the DL effect, the electron bandwidth is reduced, which leads to monotonic increases in the density of states and the static susceptibility with increasing $A_0$.
In the low-frequency limit ($\mathit{\Omega}\rightarrow 0$), the time-averaged susceptibility is exactly reduced to the equilibrium one as shown in Eq.~\eqref{eq:susceptibility_lowfreq}, since the system is equilibrated within the timescale of $\mathit{\Gamma}^{-1} \ (\ll \mathit{\Omega}^{-1})$.

In order to provide an insight into the susceptibility in the Floquet states, we performed the series expansion with respect to $A_0$ and derived the approximated expressions in Eqs.~\eqref{eq:expansion_chi_tot}--\eqref{eq:expansion_chi_inter}.
The leading-order correction to the susceptibility is proportional to $A_0^2$ and is governed by $\mathcal{A}_{\bm{k}} = \bm{v}_{\bm{k}}\bm{A}_0/\mathit{\Omega}$, where $\bm{v}_{\bm{k}}$ reflects an ``optical selection rule'' of the transitions between the Floquet bands.
It is found from the approximated expressions that the correction terms are classified into the three types of the electron-hole excitation processes: (1) excitations in the zero-photon band, $\chi^{\text{base}}$, (2) those in one of the one-photon bands, $\chi^{\text{intra}}$, and (3) those between the zero-photon band and one of the one-photon bands, $\chi^{\text{inter}}$.

In Sec.~\ref{sec:results}, we have evaluated numerically the susceptibility in Eq.~\eqref{eq:susceptibility_floquet} and verified the approximated expressions in Eqs.~\eqref{eq:expansion_chi_tot}--\eqref{eq:expansion_chi_inter}, focusing on the static limit ($\omega\rightarrow 0$).
In the two-dimensional square lattice, we demonstrated the monotonic increase in the susceptibility due to DL in the high-frequency regime and the appearance of the new peaks in the susceptibility due to the nonthermal electron distributions in the low-frequency regime where the some Floquet sidebands have the ``Fermi surfaces''.
We have also examined the one-dimensional system, where the perfect nesting is realized in equilibrium, and found the emergent peaks attributed to the inter-Floquet-band and intra-Floquet-band nesting vectors, which is illustrated in Fig.~\ref{fig:schema}.
The static susceptibilities in the real space, which describes the RKKY interaction, are presented in Figs.~\ref{fig:realspace} and \ref{fig:1d_realspace};
in the low-frequency regime, the oscillation period is modulated in accordance with the changes in $n_{\bm{k}}$ for small $A_0$.
These characteristics of the susceptibility in the Floquet states can be controlled by the amplitude, frequency, and polarization of light.

\begin{acknowledgments}
This work was supported by JSPS KAKENHI Grant No.~JP15H02100, No.~JP17H02916, No.~JP18H05208, and No.~JP18J10246.
The computation in this work has been done using the facilities of the Supercomputer Center, the Institute for Solid State Physics, the University of Tokyo.
\end{acknowledgments}

\appendix*
\section{Derivation of Eqs.~(\ref{eq:expansion_lambda})--(\ref{eq:expansion_chi_inter})}
\label{sec:derivation}

In this Appendix, we derive the series expansion of the Green functions and the susceptibility in the Floquet states with respect to $A_0$.
As mentioned in Sec.~\ref{sec:expansion}, we consider the linearly polarized light defined by
\begin{align}
\bm{A}(t) = \bm{A}_0 \sin \mathit{\Omega} t, \quad \bm{A}_0 \equiv A_0 \bm{a},
\end{align}
where $\bm{a}$ is a unit vector that represents the direction of the polarization.

First, we expand the expression of the unitary matrix $(\mathit{\Lambda}_{\bm{k}})_{mn}$ in Eq.~\eqref{eq:def_lambda} with respect to $A_0$.
Using
\begin{align}
\left. \frac{\partial \varepsilon_{0,\bm{k}}}{\partial A_0} \right\vert_{A_0=0}
&= \int_{-\pi}^{\pi} \frac{dz}{2\pi} {\left. \frac{\partial \varepsilon_{\bm{k}-A_0\bm{a}\sin z}}{\partial A_0} \right\vert_{A_0=0}} \notag \\
&= - \bm{v}_{\bm{k}} \bm{a} \int_{-\pi}^{\pi} \frac{dz}{2\pi} \sin z \notag \\
&= 0
\end{align}
and
\begin{align}
\left. \frac{\partial^2 \varepsilon_{0,\bm{k}}}{\partial A_0^2} \right\vert_{A_0=0}
&= \int_{-\pi}^{\pi} \frac{dz}{2\pi} {\left. \frac{\partial^2 \varepsilon_{\bm{k}-A_0\bm{a}\sin z}}{\partial A_0^2} \right\vert_{A_0=0}} \notag \\
&= \sum_{\alpha\beta} \tau_{\bm{k}}^{\alpha\beta} a^\alpha a^\beta \int_{-\pi}^{\pi} \frac{dz}{2\pi}\, \sin^2 z \notag \\
&= \sum_{\alpha\beta} \frac{\tau_{\bm{k}}^{\alpha\beta} a^\alpha a^\beta}{2},
\end{align}
where $\bm{v}_{\bm{k}}=\partial \varepsilon_{\bm{k}}/\partial \bm{k}$ is the group velocity and $\tau_{\bm{k}}^{\alpha\beta} = \partial^2 \varepsilon_{\bm{k}}/\partial k^\alpha \partial k^\beta$ is the energy stress tensor, we obtain the first derivative of $\mathit{\Lambda}$ as
\begin{widetext}
\begin{align}
\left.\frac{\partial (\mathit{\Lambda}_{\bm{k}})_{mn}}{\partial A_0}\right\vert_{A_0=0}
&= \int_{-\pi}^{\pi} \frac{dx}{2\pi}\, e^{i(m-n)x} {\left. \frac{\partial}{\partial A_0} \exp {\left[ \frac{1}{i\mathit{\Omega}} \int_{0}^{x} dz\, (\varepsilon_{\bm{k}-\bm{A}(z/\mathit{\Omega})} - \varepsilon_{0,\bm{k}}) \right]} \right\vert_{A_0=0}} \notag \\
&= -\frac{\bm{v}_{\bm{k}} \bm{a}}{i\mathit{\Omega}} \int_{-\pi}^{\pi} \frac{dx}{2\pi}\, e^{i(m-n)x} \int_{0}^{x} dz\, \sin z \notag \\
&= -\frac{\bm{v}_{\bm{k}}\bm{a}}{i\mathit{\Omega}} {\left[ \delta_{mn} - \frac{\delta_{m,n-1}+\delta_{m,n+1}}{2} \right]},
\end{align}
and the second derivative as
\begin{align}
\left. \frac{\partial^2 (\mathit{\Lambda}_{\bm{k}})}{\partial A_0^2} \right\vert_{A_0=0}
&= \int_{-\pi}^{\pi} \frac{dx}{2\pi}\, e^{i(m-n)x} {\left. \frac{\partial}{\partial A_0} \exp {\left[ \frac{1}{i\mathit{\Omega}} \int_0^x dz\, (\varepsilon_{\bm{k}-\bm{A}(t)}-\varepsilon_{0,\bm{k}}) \right]} \frac{1}{i\mathit{\Omega}} \int_0^x dz'\, {\left( \frac{\partial \varepsilon_{\bm{k}-\bm{A}(z'/\mathit{\Omega})}}{\partial A_0} - \frac{\partial \varepsilon_{0,\bm{k}}}{\partial A_0} \right)} \right\vert_{A_0=0}} \notag \\
&= \int_{-\pi}^{\pi} \frac{dx}{2\pi}\, e^{i(m-n)x} {\left[ {\left( \frac{-\bm{v}_{\bm{k}}\bm{a}}{i\mathit{\Omega}} \right)}^2 (1-\cos x)^2 - \frac{\tau_{\bm{k}}^{\alpha\beta} a^\alpha a^\beta}{i\mathit{\Omega}} \frac{\sin x \cos x}{2} \right]} \notag \\
&= {\left( \frac{\bm{v}_{\bm{k}}\bm{a}}{i\mathit{\Omega}} \right)}^2 {\left[ \frac{3\delta_{mn}}{2} - (\delta_{m-n,-1} + \delta_{m-n,1}) + \frac{\delta_{m-n,-2}+\delta_{m-n,2}}{4} \right]} + \frac{\tau_{\bm{k}}^{\alpha\beta} a^\alpha a^\beta}{8\mathit{\Omega}} ( \delta_{m-n,-2} - \delta_{m-n,2} ).
\end{align}
Here we adopt the summation convention for the indices $\alpha$ and $\beta$.
From these derivatives, the unitary matrix is written as
\begin{align}
(\mathit{\Lambda}_{\bm{k}})_{mn}
&= \left.(\mathit{\Lambda}_{\bm{k}})_{mn}\right\vert_{A_0=0} + {\left. \frac{\partial (\mathit{\Lambda}_{\bm{k}})_{mn}}{\partial A_0} \right\vert_{A_0=0}} A_0 + \frac{1}{2} {\left. \frac{\partial^2 (\mathit{\Lambda}_{\bm{k}})_{mn}}{\partial A_0^2} \right\vert_{A_0=0}} A_0^2 + \mathcal{O}(A_0^3) \notag \\
&= \delta_{mn} - \frac{\bm{v}_{\bm{k}}\bm{A}_0}{i\mathit{\Omega}} {\left( \delta_{mn} - \frac{\delta_{m,n-1}+\delta_{m,n+1}}{2} \right)} \notag \\
&\quad + {\left(\frac{\bm{v}_{\bm{k}}\bm{A}_0}{i\mathit{\Omega}}\right)}^2 {\left( \frac{3\delta_{mn}}{4} - \frac{\delta_{m,n-1}+\delta_{m,n+1}}{2} + \frac{\delta_{m+1,n-1}+\delta_{m-1,n+1}}{8} \right)}
+ \frac{\tau_{\bm{k}}^{\alpha\beta}A_0^\alpha A_0^\beta}{16\mathit{\Omega}} (\delta_{m+1,n-1}-\delta_{m-1,n+1})
+ \mathcal{O}(A_0^3),
\label{eq:expansion_lambda2}
\end{align}
which is Eq.~\eqref{eq:expansion_lambda}.
It is confirmed that $\mathit{\Lambda}$ in Eq.~\eqref{eq:expansion_lambda2} is unitary up to the second order in $A_0$.
Substituting Eq.~\eqref{eq:expansion_lambda2} into Eqs.~\eqref{eq:green_retarded_floquet} and \eqref{eq:green_lesser_floquet}, we obtain the expressions of the retarded Green function as
\begin{align}
(G_{\bm{k}}^{\mathrm{R}})_{mn}(\omega)
&= {\left[
1-\frac{1}{2} {\left( \frac{\bm{v}_{\bm{k}}\bm{A}_0}{\mathit{\Omega}} \right)}^2
\right]} \frac{\delta_{mn}}{\omega+n\mathit{\Omega}-\varepsilon_{0,\bm{k}}+i\eta}
+ \frac{1}{4} {\left( \frac{\bm{v}_{\bm{k}}\bm{A}_0}{\mathit{\Omega}} \right)}^2 {\left[
\frac{\delta_{mn}}{\omega+(n+1)\mathit{\Omega}-\varepsilon_{0,\bm{k}}+i\eta} + \frac{\delta_{mn}}{\omega+(n-1)\mathit{\Omega}-\varepsilon_{0,\bm{k}}+i\eta}
\right]} \notag \\
&\quad + (\delta_{m,n+1}+\delta_{m,n-1}) \frac{i\bm{v}_{\bm{k}}\bm{A}_0}{2\mathit{\Omega}} {\left(
\frac{1}{\omega+m\mathit{\Omega}-\varepsilon_{0,\bm{k}}+i\eta} - \frac{1}{\omega+n\mathit{\Omega}-\varepsilon_{0,\bm{k}}+i\eta}
\right)} \notag \\
&\quad + \delta_{m+1,n-1} {\left[
\frac{\frac{1}{4}\left(\frac{\bm{v}_{\bm{k}}\bm{A}_0}{\mathit{\Omega}}\right)^2}{\omega+(n-1)\mathit{\Omega}-\varepsilon_{0,\bm{k}}+i\eta}
- \frac{\frac{1}{8}\left(\frac{\bm{v}_{\bm{k}}\bm{A}_0}{\mathit{\Omega}}\right)^2+\frac{\tau_{\bm{k}}^{\alpha\beta}A_0^\alpha A_0^\beta}{16\mathit{\Omega}}}{\omega+m\mathit{\Omega}-\varepsilon_{0,\bm{k}}+i\eta}
- \frac{\frac{1}{8}\left(\frac{\bm{v}_{\bm{k}}\bm{A}_0}{\mathit{\Omega}}\right)^2-\frac{\tau_{\bm{k}}^{\alpha\beta}A_0^\alpha A_0^\beta}{16\mathit{\Omega}}}{\omega+n\mathit{\Omega}-\varepsilon_{0,\bm{k}}+i\eta}
\right]} \notag \\
&\quad + \delta_{m-1,n+1} {\left[
\frac{\frac{1}{4}\left(\frac{\bm{v}_{\bm{k}}\bm{A}_0}{\mathit{\Omega}}\right)^2}{\omega+(n+1)\mathit{\Omega}-\varepsilon_{0,\bm{k}}+i\eta}
- \frac{\frac{1}{8}\left(\frac{\bm{v}_{\bm{k}}\bm{A}_0}{\mathit{\Omega}}\right)^2-\frac{\tau_{\bm{k}}^{\alpha\beta}A_0^\alpha A_0^\beta}{16\mathit{\Omega}}}{\omega+m\mathit{\Omega}-\varepsilon_{0,\bm{k}}+i\eta}
- \frac{\frac{1}{8}\left(\frac{\bm{v}_{\bm{k}}\bm{A}_0}{\mathit{\Omega}}\right)^2+\frac{\tau_{\bm{k}}^{\alpha\beta}A_0^\alpha A_0^\beta}{16\mathit{\Omega}}}{\omega+n\mathit{\Omega}-\varepsilon_{0,\bm{k}}+i\eta}
\right]} + \mathcal{O}(A_0^3)
\label{eq:expansion_retarded}
\end{align}
and the lesser Green function as
\begin{align}
\frac{(G_{\bm{k}}^{<})_{mn}(\omega)}{2i\eta}
&= \delta_{mn} \Biggl[
{\left\{1-\left(\frac{\bm{v}_{\bm{k}}\bm{A}_0}{\mathit{\Omega}}\right)^2\right\}} \frac{f(\omega+n\mathit{\Omega})}{(\omega+n\mathit{\Omega}-\varepsilon_{0,\bm{k}})^2+\eta^2}
+ \frac{1}{4} {\left(\frac{\bm{v}_{\bm{k}}\bm{A}_0}{\mathit{\Omega}}\right)^2} \frac{f(\omega+(n+1)\mathit{\Omega})+f(\omega+(n-1)\mathit{\Omega})}{(\omega+n\mathit{\Omega}-\varepsilon_{0,\bm{k}})^2+\eta^2} \notag \\
&\qquad + \frac{1}{4} {\left(\frac{\bm{v}_{\bm{k}}\bm{A}_0}{\mathit{\Omega}}\right)^2} {\left\{
\frac{f(\omega+(n+1)\mathit{\Omega})}{(\omega+(n+1)\mathit{\Omega}-\varepsilon_{0,\bm{k}})^2+\eta^2}
+ \frac{f(\omega+(n-1)\mathit{\Omega})}{(\omega+(n-1)\mathit{\Omega}-\varepsilon_{0,\bm{k}})^2+\eta^2}
\right\}} \notag \\
&\qquad + \frac{1}{2} {\left(\frac{\bm{v}_{\bm{k}}\bm{A}_0}{\mathit{\Omega}}\right)^2} {\left\{
\frac{f(\omega+n\mathit{\Omega})-f(\omega+(n+1)\mathit{\Omega})}{(\omega+n\mathit{\Omega}-\varepsilon_{0,\bm{k}})(\omega+(n+1)\mathit{\Omega}-\varepsilon_{0,\bm{k}})}
+ \frac{f(\omega+n\mathit{\Omega})-f(\omega+(n-1)\mathit{\Omega})}{(\omega+n\mathit{\Omega}-\varepsilon_{0,\bm{k}})(\omega+(n-1)\mathit{\Omega}-\varepsilon_{0,\bm{k}})}
\right\}}
\Biggr] \notag \\
&\quad + (\delta_{m,n+1}+\delta_{m,n-1}) \frac{i\bm{v}_{\bm{k}}\bm{A}_0}{2\mathit{\Omega}} \notag \\
&\qquad \times {\left[
\frac{f(\omega+m\mathit{\Omega})}{(\omega+m\mathit{\Omega}-\varepsilon_{0,\bm{k}})^2+\eta^2}
- \frac{f(\omega+n\mathit{\Omega})}{(\omega+n\mathit{\Omega}-\varepsilon_{0,\bm{k}})^2+\eta^2}
- \frac{f(\omega+m\mathit{\Omega})-f(\omega+n\mathit{\Omega})}{(\omega+m\mathit{\Omega}-\varepsilon_{0,\bm{k}}+i\eta)(\omega+n\mathit{\Omega}-\varepsilon_{0,\bm{k}}-i\eta)}
\right]} \notag \\
&\quad + \delta_{m+1,n-1} \Biggl[
\frac{1}{4}\left(\frac{\bm{v}_{\bm{k}}\bm{A}_0}{\mathit{\Omega}}\right)^2 \biggl\{
\frac{f(\omega+(n-1)\mathit{\Omega})}{(\omega+(n-1)\mathit{\Omega}-\varepsilon_{0,\bm{k}})^2+\eta^2}
+ \frac{2f(\omega+(n-1)\mathit{\Omega})-f(\omega+m\mathit{\Omega})-f(\omega+n\mathit{\Omega})}{2(\omega+m\mathit{\Omega}-\varepsilon_{0,\bm{k}}+i\eta)(\omega+n\mathit{\Omega}-\varepsilon_{0,\bm{k}}-i\eta)} \notag \\
&\qquad + \frac{f(\omega+m\mathit{\Omega})-f(\omega+(n-1)\mathit{\Omega})}{(\omega+m\mathit{\Omega}-\varepsilon_{0,\bm{k}}+i\eta)(\omega+(n-1)\mathit{\Omega}-\varepsilon_{0,\bm{k}}-i\eta)}
+ \frac{f(\omega+n\mathit{\Omega})-f(\omega+(m+1)\mathit{\Omega})}{(\omega+(m+1)\mathit{\Omega}-\varepsilon_{0,\bm{k}}+i\eta)(\omega+n\mathit{\Omega}-\varepsilon_{0,\bm{k}}-i\eta)}
\biggr\} \notag \\
&\qquad - {\left\{\frac{1}{8}\left(\frac{\bm{v}_{\bm{k}}\bm{A}_0}{\mathit{\Omega}}\right)^2+\frac{\tau_{\bm{k}}^{\alpha\beta}A_0^\alpha A_0^\beta}{16\mathit{\Omega}}\right\}} \frac{f(\omega+m\mathit{\Omega})}{(\omega+m\mathit{\Omega}-\varepsilon_{0,\bm{k}})^2+\eta^2}
- {\left\{\frac{1}{8}\left(\frac{\bm{v}_{\bm{k}}\bm{A}_0}{\mathit{\Omega}}\right)^2-\frac{\tau_{\bm{k}}^{\alpha\beta}A_0^\alpha A_0^\beta}{16\mathit{\Omega}}\right\}} \frac{f(\omega+n\mathit{\Omega})}{(\omega+n\mathit{\Omega}-\varepsilon_{0,\bm{k}})^2+\eta^2} \notag \\
&\qquad + \frac{\tau_{\bm{k}}^{\alpha\beta} A_0^\alpha A_0^\beta}{16\mathit{\Omega}} \frac{f(\omega+m\mathit{\Omega})-f(\omega+n\mathit{\Omega})}{(\omega+m\mathit{\Omega}-\varepsilon_{0,\bm{k}}+i\eta)(\omega+n\mathit{\Omega}-\varepsilon_{0,\bm{k}}-i\eta)}
\Biggr] \notag \\
&\quad + \delta_{m-1,n+1} \Biggl[
\frac{1}{4}\left(\frac{\bm{v}_{\bm{k}}\bm{A}_0}{\mathit{\Omega}}\right)^2 \biggl\{
\frac{f(\omega+(n+1)\mathit{\Omega})}{(\omega+(n+1)\mathit{\Omega}-\varepsilon_{0,\bm{k}})^2+\eta^2}
+ \frac{2f(\omega+(n+1)\mathit{\Omega})-f(\omega+m\mathit{\Omega})-f(\omega+n\mathit{\Omega})}{2(\omega+m\mathit{\Omega}-\varepsilon_{0,\bm{k}}+i\eta)(\omega+n\mathit{\Omega}-\varepsilon_{0,\bm{k}}-i\eta)} \notag \\
&\qquad + \frac{f(\omega+m\mathit{\Omega})-f(\omega+(n+1)\mathit{\Omega})}{(\omega+m\mathit{\Omega}-\varepsilon_{0,\bm{k}}+i\eta)(\omega+(n+1)\mathit{\Omega}-\varepsilon_{0,\bm{k}}-i\eta)}
+ \frac{f(\omega+n\mathit{\Omega})-f(\omega+(m-1)\mathit{\Omega})}{(\omega+(m-1)\mathit{\Omega}-\varepsilon_{0,\bm{k}}+i\eta)(\omega+n\mathit{\Omega}-\varepsilon_{0,\bm{k}}-i\eta)}
\biggr\} \notag \\
&\qquad - {\left\{\frac{1}{8}\left(\frac{\bm{v}_{\bm{k}}\bm{A}_0}{\mathit{\Omega}}\right)^2-\frac{\tau_{\bm{k}}^{\alpha\beta}A_0^\alpha A_0^\beta}{16\mathit{\Omega}}\right\}} \frac{f(\omega+m\mathit{\Omega})}{(\omega+m\mathit{\Omega}-\varepsilon_{0,\bm{k}})^2+\eta^2}
- {\left\{\frac{1}{8}\left(\frac{\bm{v}_{\bm{k}}\bm{A}_0}{\mathit{\Omega}}\right)^2+\frac{\tau_{\bm{k}}^{\alpha\beta}A_0^\alpha A_0^\beta}{16\mathit{\Omega}}\right\}} \frac{f(\omega+n\mathit{\Omega})}{(\omega+n\mathit{\Omega}-\varepsilon_{0,\bm{k}})^2+\eta^2} \notag \\
&\qquad - \frac{\tau_{\bm{k}}^{\alpha\beta} A_0^\alpha A_0^\beta}{16\mathit{\Omega}} \frac{f(\omega+m\mathit{\Omega})-f(\omega+n\mathit{\Omega})}{(\omega+m\mathit{\Omega}-\varepsilon_{0,\bm{k}}+i\eta)(\omega+n\mathit{\Omega}-\varepsilon_{0,\bm{k}}-i\eta)}
\Biggr] + \mathcal{O}(A_0^3),
\label{eq:expansion_lesser}
\end{align}
where the coupling strength between the system and bath $\mathit{\Gamma}$ is replaced by the positive infinitesimal $\eta$.
The two terms in the third line in Eq.~\eqref{eq:expansion_lesser} can be neglected, since these are independent of $\eta$ and do not contribute to $G^{<}$.
The time-averaged Green functions in Eqs.~\eqref{eq:expansion_retarded_00} and \eqref{eq:expansion_lesser_00} are the $(m,n)=(0,0)$ components of the above equations.

Next, we calculate the time-averaged susceptibility, $\chi_{\bm{q}}(\omega)=(\chi_{\bm{q}})_{00}(\omega)$.
For convenience, we rewrite the susceptibility as $\chi_{\bm{q}}(\omega) = (\chi_{\bm{q}})_{00}(\omega) = \sum_{l} \chi_{\bm{q}}^{(l)}(\omega)$, where
\begin{align}
\chi_{\bm{q}}^{(l)}(\omega)
= \frac{2i}{N} \sum_{\bm{k}} \int_{-\infty}^{\infty} \frac{d\omega'}{2\pi} {\biggl[
(G_{\bm{k}+\bm{q}}^{\mathrm{R}})_{0,l}(\omega+\omega')
(G_{\bm{k}}^{<})_{l,0}(\omega')
+ (G_{\bm{k}+\bm{q}}^{<})_{0,l}(\omega+\omega')
(G_{\bm{k}}^{\mathrm{A}})_{l,0}(\omega')
\biggr]}.
\end{align}
Since the terms that contain $\delta_{m+1,n-1}$ or $\delta_{m-1,n+1}$ in Eqs.~\eqref{eq:expansion_retarded} and \eqref{eq:expansion_lesser} are proportional to $A_0^2$, we only consider the other terms which contain $\delta_{mn}$, $\delta_{m,n+1}$, and $\delta_{m,n-1}$ for $(\chi_{\bm{q}})_{00}(\omega)$, up to the second order in $A_0$.
For $l=0$, we have
\begin{align}
\chi_{\bm{q}}^{(0)}(\omega)
&= \frac{2}{N} \sum_{\bm{k}} \biggl[
\frac{f(\varepsilon_{0,\bm{k}+\bm{q}})-f(\varepsilon_{0,\bm{k}})}{\omega-(\varepsilon_{0,\bm{k}+\bm{q}}-\varepsilon_{0,\bm{k}})+2i\eta}
- \frac{\mathcal{A}_{\bm{k}+\bm{q}}^2 f(\varepsilon_{0,\bm{k}+\bm{q}}) - \mathcal{A}_{\bm{k}}^2 f(\varepsilon_{0,\bm{k}})}{\omega-(\varepsilon_{0,\bm{k}+\bm{q}}-\varepsilon_{0,\bm{k}})+2i\eta}
- \frac{1}{2} \frac{\mathcal{A}_{\bm{k}}^2 f(\varepsilon_{0,\bm{k}+\bm{q}}) - \mathcal{A}_{\bm{k}+\bm{q}}^2 f(\varepsilon_{0,\bm{k}})}{\omega-(\varepsilon_{0,\bm{k}+\bm{q}}-\varepsilon_{0,\bm{k}})+2i\eta} \notag \\
&\quad + \frac{1}{4} \frac{\mathcal{A}_{\bm{k}+\bm{q}}^2 f(\varepsilon_{0,\bm{k}+\bm{q}}+\mathit{\Omega}) - \mathcal{A}_{\bm{k}}^2 f(\varepsilon_{0,\bm{k}}+\mathit{\Omega})}{\omega-(\varepsilon_{0,\bm{k}+\bm{q}}-\varepsilon_{0,\bm{k}})+2i\eta}
+ \frac{1}{4} \frac{\mathcal{A}_{\bm{k}+\bm{q}}^2 f(\varepsilon_{0,\bm{k}+\bm{q}}-\mathit{\Omega}) - \mathcal{A}_{\bm{k}}^2 f(\varepsilon_{0,\bm{k}}-\mathit{\Omega})}{\omega-(\varepsilon_{0,\bm{k}+\bm{q}}-\varepsilon_{0,\bm{k}})+2i\eta} \notag \\
&\quad + \frac{\mathcal{A}_{\bm{k}+\bm{q}}^2+\mathcal{A}_{\bm{k}}^2}{4} {\left\{
\frac{f(\varepsilon_{0,\bm{k}+\bm{q}})-f(\varepsilon_{0,\bm{k}})}{\omega+\mathit{\Omega}-(\varepsilon_{0,\bm{k}+\bm{q}}-\varepsilon_{0,\bm{k}})+2i\eta}
+ \frac{f(\varepsilon_{0,\bm{k}+\bm{q}})-f(\varepsilon_{0,\bm{k}})}{\omega-\mathit{\Omega}-(\varepsilon_{0,\bm{k}+\bm{q}}-\varepsilon_{0,\bm{k}})+2i\eta}
\right\}}
\biggr] + \mathcal{O}(A_0^3),
\label{eq:derivation_0}
\end{align}
where we use $\lim_{\eta\rightarrow 0} \eta/(z^2+\eta^2) = \pi \delta(z)$ and define $\mathcal{A}_{\bm{k}}=\bm{v}_{\bm{k}}\bm{A}_0/\mathit{\Omega}$.
This is reduced to Eq.~\eqref{eq:susceptibility_highfreq} in the limit of $\mathit{\Omega}\rightarrow \infty$ or $A_0\rightarrow 0$.
Similarly to $\chi_{\bm{q}}^{(0)}(\omega)$, for $l=\pm 1$, we obtain
\begin{align}
\chi_{\bm{q}}^{(l)}(\omega)
&= \frac{2}{N} \sum_{\bm{k}} \frac{\mathcal{A}_{\bm{k}+\bm{q}} \mathcal{A}_{\bm{k}}}{4} \biggl[
\biggl\{
\frac{2\{f(\varepsilon_{0,\bm{k}+\bm{q}})-f(\varepsilon_{0,\bm{k}})\}}{\omega-(\varepsilon_{0,\bm{k}+\bm{q}}-\varepsilon_{0,\bm{k}})+2i\eta} 
 - \frac{f(\varepsilon_{0,\bm{k}+\bm{q}})-f(\varepsilon_{0,\bm{k}})}{\omega+l\mathit{\Omega}-(\varepsilon_{0,\bm{k}+\bm{q}}-\varepsilon_{0,\bm{k}})+2i\eta}
- \frac{f(\varepsilon_{0,\bm{k}+\bm{q}})-f(\varepsilon_{0,\bm{k}})}{\omega-l\mathit{\Omega}-(\varepsilon_{0,\bm{k}+\bm{q}}-\varepsilon_{0,\bm{k}})+2i\eta}
\biggr\} \notag \\
&\quad + \int_{-\infty}^{\infty} \frac{d\omega'}{2\pi} \frac{i}{\omega+\omega'-\varepsilon_{0,\bm{k}+\bm{q}}+i\eta} \biggl\{
\frac{2i\eta \{f(\omega')-f(\omega'+l\mathit{\Omega})\}}{(\omega'+l\mathit{\Omega}-\varepsilon_{0,\bm{k}}+i\eta)(\omega'-\varepsilon_{0,\bm{k}}-i\eta)}
+ \frac{2i\eta \{f(\omega')-f(\omega'-l\mathit{\Omega})\}}{(\omega'-\varepsilon_{0,\bm{k}}+i\eta)(\omega'-l\mathit{\Omega}-\varepsilon_{0,\bm{k}}-i\eta)}
\biggr\} \notag \\
&\quad + \int_{-\infty}^{\infty} \frac{d\omega'}{2\pi} \biggl\{
\frac{2i\eta \{f(\omega+\omega')-f(\omega+\omega'-l\mathit{\Omega})\}}{(\omega+\omega'-l\mathit{\Omega}-\varepsilon_{0,\bm{k}+\bm{q}}+i\eta)(\omega+\omega'-\varepsilon_{0,\bm{k}+\bm{q}}-i\eta)} \notag \\
&\qquad + \frac{2i\eta \{f(\omega+\omega')-f(\omega+\omega'+l\mathit{\Omega})\}}{(\omega+\omega'-\varepsilon_{0,\bm{k}+\bm{q}}+i\eta)(\omega+\omega'+l\mathit{\Omega}-\varepsilon_{0,\bm{k}+\bm{q}}-i\eta)}
\biggr\}
\frac{i}{\omega'-\varepsilon_{0,\bm{k}}-i\eta}
\biggr] + \mathcal{O}(A_0^3),
\end{align}
and then
\begin{align}
\sum_{l=\pm 1} \chi_{\bm{q}}^{(l)}(\omega)
&= \frac{2}{N} \sum_{\bm{k}} \frac{\mathcal{A}_{\bm{k}+\bm{q}} \mathcal{A}_{\bm{k}}}{2} \biggl[
\biggl\{
\frac{2\{f(\varepsilon_{0,\bm{k}+\bm{q}})-f(\varepsilon_{0,\bm{k}})\}}{\omega-(\varepsilon_{0,\bm{k}+\bm{q}}-\varepsilon_{0,\bm{k}})+2i\eta} 
 - \frac{f(\varepsilon_{0,\bm{k}+\bm{q}})-f(\varepsilon_{0,\bm{k}})}{\omega+\mathit{\Omega}-(\varepsilon_{0,\bm{k}+\bm{q}}-\varepsilon_{0,\bm{k}})+2i\eta}
- \frac{f(\varepsilon_{0,\bm{k}+\bm{q}})-f(\varepsilon_{0,\bm{k}})}{\omega-\mathit{\Omega}-(\varepsilon_{0,\bm{k}+\bm{q}}-\varepsilon_{0,\bm{k}})+2i\eta}
\biggr\} \notag \\
&\qquad + \int_{-\infty}^{\infty} \frac{d\omega'}{2\pi} \biggl\{
\frac{i}{\omega+\omega'-\varepsilon_{0,\bm{k}+\bm{q}}+i\eta} \biggl(
\frac{2i\eta \{f(\omega')-f(\omega'+\mathit{\Omega})\}}{(\omega'-\varepsilon_{0,\bm{k}})(\omega'+\mathit{\Omega}-\varepsilon_{0,\bm{k}})}
+ \frac{2i\eta \{f(\omega')-f(\omega'-\mathit{\Omega})\}}{(\omega'-\varepsilon_{0,\bm{k}})(\omega'-\mathit{\Omega}-\varepsilon_{0,\bm{k}})}
\biggr) \notag \\
&\qquad + \biggl(
\frac{2i\eta \{f(\omega+\omega')-f(\omega+\omega'+\mathit{\Omega})\}}{(\omega+\omega'-\varepsilon_{0,\bm{k}+\bm{q}})(\omega+\omega'+\mathit{\Omega}-\varepsilon_{0,\bm{k}+\bm{q}})}
+ \frac{2i\eta \{f(\omega+\omega')-f(\omega+\omega'-\mathit{\Omega})\}}{(\omega+\omega'-\varepsilon_{0,\bm{k}+\bm{q}})(\omega+\omega'-\mathit{\Omega}-\varepsilon_{0,\bm{k}+\bm{q}})}
\biggr)
\frac{i}{\omega'-\varepsilon_{0,\bm{k}}-i\eta}
\biggr\}
\biggr] \notag \\
&\quad + \mathcal{O}(A_0^3),
\label{eq:derivation_1}
\end{align}
\end{widetext}
where the second and third lines in Eq.~\eqref{eq:derivation_1} vanish due to $\eta$ in the numerators.
After substituting Eqs.~\eqref{eq:derivation_0} and \eqref{eq:derivation_1} into $\chi_{\bm{q}}(\omega) = \sum_{l} \chi_{\bm{q}}^{(l)}(\omega)$, we have Eqs.~\eqref{eq:expansion_chi_tot}--\eqref{eq:expansion_chi_inter}.

The time-averaged energy band $\varepsilon_{0,\bm{k}}$ depends on $A_0$, whose leading-order correction to $\varepsilon_{\bm{k}}$ is also $\mathcal{O}(A_0^2)$.
It is found from the above derivation that a dimensionless parameter which governs the series expansion of $\mathit{\Lambda}$ is $\mathcal{A}_{\bm{k}}=\bm{v}_{\bm{k}}\bm{A}_0/\mathit{\Omega}$ rather than $A_0$, while the counterpart of $\varepsilon_{0,\bm{k}}$ is $A_0$.
Therefore, we do not expand $\varepsilon_{0,\bm{k}}$ with respect to $A_0$.

\bibliography{reference}

\begin{thebibliography}{64}%
\makeatletter
\providecommand \@ifxundefined [1]{%
 \@ifx{#1\undefined}
}%
\providecommand \@ifnum [1]{%
 \ifnum #1\expandafter \@firstoftwo
 \else \expandafter \@secondoftwo
 \fi
}%
\providecommand \@ifx [1]{%
 \ifx #1\expandafter \@firstoftwo
 \else \expandafter \@secondoftwo
 \fi
}%
\providecommand \natexlab [1]{#1}%
\providecommand \enquote  [1]{``#1''}%
\providecommand \bibnamefont  [1]{#1}%
\providecommand \bibfnamefont [1]{#1}%
\providecommand \citenamefont [1]{#1}%
\providecommand \href@noop [0]{\@secondoftwo}%
\providecommand \href [0]{\begingroup \@sanitize@url \@href}%
\providecommand \@href[1]{\@@startlink{#1}\@@href}%
\providecommand \@@href[1]{\endgroup#1\@@endlink}%
\providecommand \@sanitize@url [0]{\catcode `\\12\catcode `\$12\catcode
  `\&12\catcode `\#12\catcode `\^12\catcode `\_12\catcode `\%12\relax}%
\providecommand \@@startlink[1]{}%
\providecommand \@@endlink[0]{}%
\providecommand \url  [0]{\begingroup\@sanitize@url \@url }%
\providecommand \@url [1]{\endgroup\@href {#1}{\urlprefix }}%
\providecommand \urlprefix  [0]{URL }%
\providecommand \Eprint [0]{\href }%
\providecommand \doibase [0]{http://dx.doi.org/}%
\providecommand \selectlanguage [0]{\@gobble}%
\providecommand \bibinfo  [0]{\@secondoftwo}%
\providecommand \bibfield  [0]{\@secondoftwo}%
\providecommand \translation [1]{[#1]}%
\providecommand \BibitemOpen [0]{}%
\providecommand \bibitemStop [0]{}%
\providecommand \bibitemNoStop [0]{.\EOS\space}%
\providecommand \EOS [0]{\spacefactor3000\relax}%
\providecommand \BibitemShut  [1]{\csname bibitem#1\endcsname}%
\let\auto@bib@innerbib\@empty
\bibitem [{\citenamefont {Nasu}(2004)}]{Nasu2004}%
  \BibitemOpen
  \bibinfo {editor} {\bibfnamefont {K.}~\bibnamefont {Nasu}},\ ed.,\ \href@noop
  {} {\emph {\bibinfo {title} {Photoinduced Phase Transitions}}}\ (\bibinfo
  {publisher} {World Scientific},\ \bibinfo {address} {Singapore},\ \bibinfo
  {year} {2004})\BibitemShut {NoStop}%
\bibitem [{\citenamefont {Tokura}(2006)}]{Tokura2006}%
  \BibitemOpen
  \bibfield  {author} {\bibinfo {author} {\bibfnamefont {Y.}~\bibnamefont
  {Tokura}},\ }\href {\doibase 10.1143/JPSJ.75.011001} {\bibfield  {journal}
  {\bibinfo  {journal} {J. Phys. Soc. Jpn.}\ }\textbf {\bibinfo {volume}
  {75}},\ \bibinfo {pages} {011001} (\bibinfo {year} {2006})}\BibitemShut
  {NoStop}%
\bibitem [{\citenamefont {Basov}\ \emph {et~al.}(2011)\citenamefont {Basov},
  \citenamefont {Averitt}, \citenamefont {van~der Marel}, \citenamefont
  {Dressel},\ and\ \citenamefont {Haule}}]{Basov2011}%
  \BibitemOpen
  \bibfield  {author} {\bibinfo {author} {\bibfnamefont {D.~N.}\ \bibnamefont
  {Basov}}, \bibinfo {author} {\bibfnamefont {R.~D.}\ \bibnamefont {Averitt}},
  \bibinfo {author} {\bibfnamefont {D.}~\bibnamefont {van~der Marel}}, \bibinfo
  {author} {\bibfnamefont {M.}~\bibnamefont {Dressel}}, \ and\ \bibinfo
  {author} {\bibfnamefont {K.}~\bibnamefont {Haule}},\ }\href {\doibase
  10.1103/RevModPhys.83.471} {\bibfield  {journal} {\bibinfo  {journal} {Rev.
  Mod. Phys.}\ }\textbf {\bibinfo {volume} {83}},\ \bibinfo {pages} {471}
  (\bibinfo {year} {2011})}\BibitemShut {NoStop}%
\bibitem [{\citenamefont {Kirilyuk}\ \emph {et~al.}(2010)\citenamefont
  {Kirilyuk}, \citenamefont {Kimel},\ and\ \citenamefont
  {Rasing}}]{Kirilyuk2010}%
  \BibitemOpen
  \bibfield  {author} {\bibinfo {author} {\bibfnamefont {A.}~\bibnamefont
  {Kirilyuk}}, \bibinfo {author} {\bibfnamefont {A.~V.}\ \bibnamefont {Kimel}},
  \ and\ \bibinfo {author} {\bibfnamefont {T.}~\bibnamefont {Rasing}},\ }\href
  {\doibase 10.1103/RevModPhys.82.2731} {\bibfield  {journal} {\bibinfo
  {journal} {Rev. Mod. Phys.}\ }\textbf {\bibinfo {volume} {82}},\ \bibinfo
  {pages} {2731} (\bibinfo {year} {2010})}\BibitemShut {NoStop}%
\bibitem [{\citenamefont {Mentink}(2017)}]{Mentink2017}%
  \BibitemOpen
  \bibfield  {author} {\bibinfo {author} {\bibfnamefont {J.~H.}\ \bibnamefont
  {Mentink}},\ }\href {\doibase 10.1088/1361-648X/aa8abf} {\bibfield  {journal}
  {\bibinfo  {journal} {J. Phys. Condens. Matter}\ }\textbf {\bibinfo {volume}
  {29}},\ \bibinfo {pages} {453001} (\bibinfo {year} {2017})}\BibitemShut
  {NoStop}%
\bibitem [{\citenamefont {Kampfrath}\ \emph {et~al.}(2013)\citenamefont
  {Kampfrath}, \citenamefont {Tanaka},\ and\ \citenamefont
  {Nelson}}]{Kampfrath2013}%
  \BibitemOpen
  \bibfield  {author} {\bibinfo {author} {\bibfnamefont {T.}~\bibnamefont
  {Kampfrath}}, \bibinfo {author} {\bibfnamefont {K.}~\bibnamefont {Tanaka}}, \
  and\ \bibinfo {author} {\bibfnamefont {K.~A.}\ \bibnamefont {Nelson}},\
  }\href {\doibase 10.1038/nphoton.2013.184} {\bibfield  {journal} {\bibinfo
  {journal} {Nat. Photonics}\ }\textbf {\bibinfo {volume} {7}},\ \bibinfo
  {pages} {680} (\bibinfo {year} {2013})}\BibitemShut {NoStop}%
\bibitem [{\citenamefont {Miyamoto}\ \emph {et~al.}(2018)\citenamefont
  {Miyamoto}, \citenamefont {Yamakawa}, \citenamefont {Morimoto},\ and\
  \citenamefont {Okamoto}}]{Miyamoto2018}%
  \BibitemOpen
  \bibfield  {author} {\bibinfo {author} {\bibfnamefont {T.}~\bibnamefont
  {Miyamoto}}, \bibinfo {author} {\bibfnamefont {H.}~\bibnamefont {Yamakawa}},
  \bibinfo {author} {\bibfnamefont {T.}~\bibnamefont {Morimoto}}, \ and\
  \bibinfo {author} {\bibfnamefont {H.}~\bibnamefont {Okamoto}},\ }\href
  {\doibase 10.1088/1361-6455/aad023} {\bibfield  {journal} {\bibinfo
  {journal} {J. Phys. B}\ }\textbf {\bibinfo {volume} {51}},\ \bibinfo {pages}
  {162001} (\bibinfo {year} {2018})}\BibitemShut {NoStop}%
\bibitem [{\citenamefont {Kawakami}\ \emph {et~al.}(2018)\citenamefont
  {Kawakami}, \citenamefont {Itoh}, \citenamefont {Yonemitsu},\ and\
  \citenamefont {Iwai}}]{Kawakami2018}%
  \BibitemOpen
  \bibfield  {author} {\bibinfo {author} {\bibfnamefont {Y.}~\bibnamefont
  {Kawakami}}, \bibinfo {author} {\bibfnamefont {H.}~\bibnamefont {Itoh}},
  \bibinfo {author} {\bibfnamefont {K.}~\bibnamefont {Yonemitsu}}, \ and\
  \bibinfo {author} {\bibfnamefont {S.}~\bibnamefont {Iwai}},\ }\href {\doibase
  10.1088/1361-6455/aad40a} {\bibfield  {journal} {\bibinfo  {journal} {J.
  Phys. B}\ }\textbf {\bibinfo {volume} {51}},\ \bibinfo {pages} {174005}
  (\bibinfo {year} {2018})}\BibitemShut {NoStop}%
\bibitem [{\citenamefont {Shirley}(1965)}]{Shirley1965}%
  \BibitemOpen
  \bibfield  {author} {\bibinfo {author} {\bibfnamefont {J.}~\bibnamefont
  {Shirley}},\ }\href {\doibase 10.1103/PhysRev.138.B979} {\bibfield  {journal}
  {\bibinfo  {journal} {Phys. Rev.}\ }\textbf {\bibinfo {volume} {138}},\
  \bibinfo {pages} {B979} (\bibinfo {year} {1965})}\BibitemShut {NoStop}%
\bibitem [{\citenamefont {Sambe}(1973)}]{Sambe1973}%
  \BibitemOpen
  \bibfield  {author} {\bibinfo {author} {\bibfnamefont {H.}~\bibnamefont
  {Sambe}},\ }\href {\doibase 10.1103/PhysRevA.7.2203} {\bibfield  {journal}
  {\bibinfo  {journal} {Phys. Rev. A}\ }\textbf {\bibinfo {volume} {7}},\
  \bibinfo {pages} {2203} (\bibinfo {year} {1973})}\BibitemShut {NoStop}%
\bibitem [{\citenamefont {Aoki}\ \emph {et~al.}(2014)\citenamefont {Aoki},
  \citenamefont {Tsuji}, \citenamefont {Eckstein}, \citenamefont {Kollar},
  \citenamefont {Oka},\ and\ \citenamefont {Werner}}]{Aoki2014}%
  \BibitemOpen
  \bibfield  {author} {\bibinfo {author} {\bibfnamefont {H.}~\bibnamefont
  {Aoki}}, \bibinfo {author} {\bibfnamefont {N.}~\bibnamefont {Tsuji}},
  \bibinfo {author} {\bibfnamefont {M.}~\bibnamefont {Eckstein}}, \bibinfo
  {author} {\bibfnamefont {M.}~\bibnamefont {Kollar}}, \bibinfo {author}
  {\bibfnamefont {T.}~\bibnamefont {Oka}}, \ and\ \bibinfo {author}
  {\bibfnamefont {P.}~\bibnamefont {Werner}},\ }\href {\doibase
  10.1103/RevModPhys.86.779} {\bibfield  {journal} {\bibinfo  {journal} {Rev.
  Mod. Phys.}\ }\textbf {\bibinfo {volume} {86}},\ \bibinfo {pages} {779}
  (\bibinfo {year} {2014})}\BibitemShut {NoStop}%
\bibitem [{\citenamefont {Wang}\ \emph {et~al.}(2013)\citenamefont {Wang},
  \citenamefont {Steinberg}, \citenamefont {Jarillo-Herrero},\ and\
  \citenamefont {Gedik}}]{Wang2013}%
  \BibitemOpen
  \bibfield  {author} {\bibinfo {author} {\bibfnamefont {Y.~H.}\ \bibnamefont
  {Wang}}, \bibinfo {author} {\bibfnamefont {H.}~\bibnamefont {Steinberg}},
  \bibinfo {author} {\bibfnamefont {P.}~\bibnamefont {Jarillo-Herrero}}, \ and\
  \bibinfo {author} {\bibfnamefont {N.}~\bibnamefont {Gedik}},\ }\href
  {\doibase 10.1126/science.1239834} {\bibfield  {journal} {\bibinfo  {journal}
  {Science}\ }\textbf {\bibinfo {volume} {342}},\ \bibinfo {pages} {453}
  (\bibinfo {year} {2013})}\BibitemShut {NoStop}%
\bibitem [{\citenamefont {Mahmood}\ \emph {et~al.}(2016)\citenamefont
  {Mahmood}, \citenamefont {Chan}, \citenamefont {Alpichshev}, \citenamefont
  {Gardner}, \citenamefont {Lee}, \citenamefont {Lee},\ and\ \citenamefont
  {Gedik}}]{Mahmood2016}%
  \BibitemOpen
  \bibfield  {author} {\bibinfo {author} {\bibfnamefont {F.}~\bibnamefont
  {Mahmood}}, \bibinfo {author} {\bibfnamefont {C.-K.}\ \bibnamefont {Chan}},
  \bibinfo {author} {\bibfnamefont {Z.}~\bibnamefont {Alpichshev}}, \bibinfo
  {author} {\bibfnamefont {D.}~\bibnamefont {Gardner}}, \bibinfo {author}
  {\bibfnamefont {Y.}~\bibnamefont {Lee}}, \bibinfo {author} {\bibfnamefont
  {P.~A.}\ \bibnamefont {Lee}}, \ and\ \bibinfo {author} {\bibfnamefont
  {N.}~\bibnamefont {Gedik}},\ }\href {\doibase 10.1038/nphys3609} {\bibfield
  {journal} {\bibinfo  {journal} {Nat. Phys.}\ }\textbf {\bibinfo {volume}
  {12}},\ \bibinfo {pages} {306} (\bibinfo {year} {2016})}\BibitemShut
  {NoStop}%
\bibitem [{\citenamefont {Bukov}\ \emph {et~al.}(2015)\citenamefont {Bukov},
  \citenamefont {D'Alessio},\ and\ \citenamefont {Polkovnikov}}]{Bukov2015}%
  \BibitemOpen
  \bibfield  {author} {\bibinfo {author} {\bibfnamefont {M.}~\bibnamefont
  {Bukov}}, \bibinfo {author} {\bibfnamefont {L.}~\bibnamefont {D'Alessio}}, \
  and\ \bibinfo {author} {\bibfnamefont {A.}~\bibnamefont {Polkovnikov}},\
  }\href {\doibase 10.1080/00018732.2015.1055918} {\bibfield  {journal}
  {\bibinfo  {journal} {Adv. Phys.}\ }\textbf {\bibinfo {volume} {64}},\
  \bibinfo {pages} {139} (\bibinfo {year} {2015})}\BibitemShut {NoStop}%
\bibitem [{\citenamefont {Eckardt}(2017)}]{Eckardt2017}%
  \BibitemOpen
  \bibfield  {author} {\bibinfo {author} {\bibfnamefont {A.}~\bibnamefont
  {Eckardt}},\ }\href {\doibase 10.1103/RevModPhys.89.011004} {\bibfield
  {journal} {\bibinfo  {journal} {Rev. Mod. Phys.}\ }\textbf {\bibinfo {volume}
  {89}},\ \bibinfo {pages} {011004} (\bibinfo {year} {2017})}\BibitemShut
  {NoStop}%
\bibitem [{\citenamefont {Oka}\ and\ \citenamefont {Kitamura}()}]{Oka2018}%
  \BibitemOpen
  \bibfield  {author} {\bibinfo {author} {\bibfnamefont {T.}~\bibnamefont
  {Oka}}\ and\ \bibinfo {author} {\bibfnamefont {S.}~\bibnamefont {Kitamura}},\
  }\href@noop {} {}\Eprint {http://arxiv.org/abs/1804.03212} {arXiv:1804.03212}
  \BibitemShut {NoStop}%
\bibitem [{\citenamefont {Eckstein}\ and\ \citenamefont
  {Kollar}(2008)}]{Eckstein2008}%
  \BibitemOpen
  \bibfield  {author} {\bibinfo {author} {\bibfnamefont {M.}~\bibnamefont
  {Eckstein}}\ and\ \bibinfo {author} {\bibfnamefont {M.}~\bibnamefont
  {Kollar}},\ }\href {\doibase 10.1103/PhysRevB.78.205119} {\bibfield
  {journal} {\bibinfo  {journal} {Phys. Rev. B}\ }\textbf {\bibinfo {volume}
  {78}},\ \bibinfo {pages} {205119} (\bibinfo {year} {2008})}\BibitemShut
  {NoStop}%
\bibitem [{\citenamefont {Tsuji}\ \emph {et~al.}(2009)\citenamefont {Tsuji},
  \citenamefont {Oka},\ and\ \citenamefont {Aoki}}]{Tsuji2009}%
  \BibitemOpen
  \bibfield  {author} {\bibinfo {author} {\bibfnamefont {N.}~\bibnamefont
  {Tsuji}}, \bibinfo {author} {\bibfnamefont {T.}~\bibnamefont {Oka}}, \ and\
  \bibinfo {author} {\bibfnamefont {H.}~\bibnamefont {Aoki}},\ }\href {\doibase
  10.1103/PhysRevLett.103.047403} {\bibfield  {journal} {\bibinfo  {journal}
  {Phys. Rev. Lett.}\ }\textbf {\bibinfo {volume} {103}},\ \bibinfo {pages}
  {047403} (\bibinfo {year} {2009})}\BibitemShut {NoStop}%
\bibitem [{\citenamefont {Tsuji}\ and\ \citenamefont {Aoki}(2015)}]{Tsuji2015}%
  \BibitemOpen
  \bibfield  {author} {\bibinfo {author} {\bibfnamefont {N.}~\bibnamefont
  {Tsuji}}\ and\ \bibinfo {author} {\bibfnamefont {H.}~\bibnamefont {Aoki}},\
  }\href {\doibase 10.1103/PhysRevB.92.064508} {\bibfield  {journal} {\bibinfo
  {journal} {Phys. Rev. B}\ }\textbf {\bibinfo {volume} {92}},\ \bibinfo
  {pages} {064508} (\bibinfo {year} {2015})}\BibitemShut {NoStop}%
\bibitem [{\citenamefont {Perfetto}\ and\ \citenamefont
  {Stefanucci}(2015)}]{Perfetto2015b}%
  \BibitemOpen
  \bibfield  {author} {\bibinfo {author} {\bibfnamefont {E.}~\bibnamefont
  {Perfetto}}\ and\ \bibinfo {author} {\bibfnamefont {G.}~\bibnamefont
  {Stefanucci}},\ }\href {\doibase 10.1103/PhysRevA.91.033416} {\bibfield
  {journal} {\bibinfo  {journal} {Phys. Rev. A}\ }\textbf {\bibinfo {volume}
  {91}},\ \bibinfo {pages} {033416} (\bibinfo {year} {2015})}\BibitemShut
  {NoStop}%
\bibitem [{\citenamefont {Perfetto}\ \emph {et~al.}(2015)\citenamefont
  {Perfetto}, \citenamefont {Sangalli}, \citenamefont {Marini},\ and\
  \citenamefont {Stefanucci}}]{Perfetto2015}%
  \BibitemOpen
  \bibfield  {author} {\bibinfo {author} {\bibfnamefont {E.}~\bibnamefont
  {Perfetto}}, \bibinfo {author} {\bibfnamefont {D.}~\bibnamefont {Sangalli}},
  \bibinfo {author} {\bibfnamefont {A.}~\bibnamefont {Marini}}, \ and\ \bibinfo
  {author} {\bibfnamefont {G.}~\bibnamefont {Stefanucci}},\ }\href {\doibase
  10.1103/PhysRevB.92.205304} {\bibfield  {journal} {\bibinfo  {journal} {Phys.
  Rev. B}\ }\textbf {\bibinfo {volume} {92}},\ \bibinfo {pages} {205304}
  (\bibinfo {year} {2015})}\BibitemShut {NoStop}%
\bibitem [{\citenamefont {Matsueda}\ and\ \citenamefont
  {Ishihara}(2007)}]{Matsueda2007}%
  \BibitemOpen
  \bibfield  {author} {\bibinfo {author} {\bibfnamefont {H.}~\bibnamefont
  {Matsueda}}\ and\ \bibinfo {author} {\bibfnamefont {S.}~\bibnamefont
  {Ishihara}},\ }\href {\doibase 10.1143/JPSJ.76.083703} {\bibfield  {journal}
  {\bibinfo  {journal} {J. Phys. Soc. Jpn.}\ }\textbf {\bibinfo {volume}
  {76}},\ \bibinfo {pages} {083703} (\bibinfo {year} {2007})}\BibitemShut
  {NoStop}%
\bibitem [{\citenamefont {Kanamori}\ \emph {et~al.}(2011)\citenamefont
  {Kanamori}, \citenamefont {Matsueda},\ and\ \citenamefont
  {Ishihara}}]{Kanamori2011}%
  \BibitemOpen
  \bibfield  {author} {\bibinfo {author} {\bibfnamefont {Y.}~\bibnamefont
  {Kanamori}}, \bibinfo {author} {\bibfnamefont {H.}~\bibnamefont {Matsueda}},
  \ and\ \bibinfo {author} {\bibfnamefont {S.}~\bibnamefont {Ishihara}},\
  }\href {\doibase 10.1103/PhysRevLett.107.167403} {\bibfield  {journal}
  {\bibinfo  {journal} {Phys. Rev. Lett.}\ }\textbf {\bibinfo {volume} {107}},\
  \bibinfo {pages} {167403} (\bibinfo {year} {2011})}\BibitemShut {NoStop}%
\bibitem [{\citenamefont {Iyoda}\ and\ \citenamefont
  {Ishihara}(2014)}]{Iyoda2014}%
  \BibitemOpen
  \bibfield  {author} {\bibinfo {author} {\bibfnamefont {E.}~\bibnamefont
  {Iyoda}}\ and\ \bibinfo {author} {\bibfnamefont {S.}~\bibnamefont
  {Ishihara}},\ }\href {\doibase 10.1103/PhysRevB.89.125126} {\bibfield
  {journal} {\bibinfo  {journal} {Phys. Rev. B}\ }\textbf {\bibinfo {volume}
  {89}},\ \bibinfo {pages} {125126} (\bibinfo {year} {2014})}\BibitemShut
  {NoStop}%
\bibitem [{\citenamefont {Lenar{\v{c}}i{\v{c}}}\ \emph
  {et~al.}(2014)\citenamefont {Lenar{\v{c}}i{\v{c}}}, \citenamefont
  {Gole{\v{z}}}, \citenamefont {Bon{\v{c}}a},\ and\ \citenamefont
  {Prelov{\v{s}}ek}}]{Lenarcic2014}%
  \BibitemOpen
  \bibfield  {author} {\bibinfo {author} {\bibfnamefont {Z.}~\bibnamefont
  {Lenar{\v{c}}i{\v{c}}}}, \bibinfo {author} {\bibfnamefont {D.}~\bibnamefont
  {Gole{\v{z}}}}, \bibinfo {author} {\bibfnamefont {J.}~\bibnamefont
  {Bon{\v{c}}a}}, \ and\ \bibinfo {author} {\bibfnamefont {P.}~\bibnamefont
  {Prelov{\v{s}}ek}},\ }\href {\doibase 10.1103/PhysRevB.89.125123} {\bibfield
  {journal} {\bibinfo  {journal} {Phys. Rev. B}\ }\textbf {\bibinfo {volume}
  {89}},\ \bibinfo {pages} {125123} (\bibinfo {year} {2014})}\BibitemShut
  {NoStop}%
\bibitem [{\citenamefont {Kogoj}\ \emph {et~al.}(2016)\citenamefont {Kogoj},
  \citenamefont {Vidmar}, \citenamefont {Mierzejewski}, \citenamefont
  {Trugman},\ and\ \citenamefont {Bon{\v{c}}a}}]{Kogoj2016}%
  \BibitemOpen
  \bibfield  {author} {\bibinfo {author} {\bibfnamefont {J.}~\bibnamefont
  {Kogoj}}, \bibinfo {author} {\bibfnamefont {L.}~\bibnamefont {Vidmar}},
  \bibinfo {author} {\bibfnamefont {M.}~\bibnamefont {Mierzejewski}}, \bibinfo
  {author} {\bibfnamefont {S.~A.}\ \bibnamefont {Trugman}}, \ and\ \bibinfo
  {author} {\bibfnamefont {J.}~\bibnamefont {Bon{\v{c}}a}},\ }\href {\doibase
  10.1103/PhysRevB.94.014304} {\bibfield  {journal} {\bibinfo  {journal} {Phys.
  Rev. B}\ }\textbf {\bibinfo {volume} {94}},\ \bibinfo {pages} {014304}
  (\bibinfo {year} {2016})}\BibitemShut {NoStop}%
\bibitem [{\citenamefont {Shao}\ \emph {et~al.}(2016)\citenamefont {Shao},
  \citenamefont {Tohyama}, \citenamefont {Luo},\ and\ \citenamefont
  {Lu}}]{Shao2016}%
  \BibitemOpen
  \bibfield  {author} {\bibinfo {author} {\bibfnamefont {C.}~\bibnamefont
  {Shao}}, \bibinfo {author} {\bibfnamefont {T.}~\bibnamefont {Tohyama}},
  \bibinfo {author} {\bibfnamefont {H.-G.}\ \bibnamefont {Luo}}, \ and\
  \bibinfo {author} {\bibfnamefont {H.}~\bibnamefont {Lu}},\ }\href {\doibase
  10.1103/PhysRevB.93.195144} {\bibfield  {journal} {\bibinfo  {journal} {Phys.
  Rev. B}\ }\textbf {\bibinfo {volume} {93}},\ \bibinfo {pages} {195144}
  (\bibinfo {year} {2016})}\BibitemShut {NoStop}%
\bibitem [{\citenamefont {Shinjo}\ and\ \citenamefont
  {Tohyama}(2017)}]{Shinjo2017}%
  \BibitemOpen
  \bibfield  {author} {\bibinfo {author} {\bibfnamefont {K.}~\bibnamefont
  {Shinjo}}\ and\ \bibinfo {author} {\bibfnamefont {T.}~\bibnamefont
  {Tohyama}},\ }\href {\doibase 10.1103/PhysRevB.96.195141} {\bibfield
  {journal} {\bibinfo  {journal} {Phys. Rev. B}\ }\textbf {\bibinfo {volume}
  {96}},\ \bibinfo {pages} {195141} (\bibinfo {year} {2017})}\BibitemShut
  {NoStop}%
\bibitem [{\citenamefont {Ono}\ and\ \citenamefont {Ishihara}()}]{Ono2018}%
  \BibitemOpen
  \bibfield  {author} {\bibinfo {author} {\bibfnamefont {A.}~\bibnamefont
  {Ono}}\ and\ \bibinfo {author} {\bibfnamefont {S.}~\bibnamefont {Ishihara}},\
  }\href@noop {} {}\Eprint {http://arxiv.org/abs/1809.07132} {arXiv:1809.07132}
  \BibitemShut {NoStop}%
\bibitem [{\citenamefont {Bittner}\ \emph {et~al.}(2018)\citenamefont
  {Bittner}, \citenamefont {Gole{\v{z}}}, \citenamefont {Strand}, \citenamefont
  {Eckstein},\ and\ \citenamefont {Werner}}]{Bittner2018}%
  \BibitemOpen
  \bibfield  {author} {\bibinfo {author} {\bibfnamefont {N.}~\bibnamefont
  {Bittner}}, \bibinfo {author} {\bibfnamefont {D.}~\bibnamefont
  {Gole{\v{z}}}}, \bibinfo {author} {\bibfnamefont {H.~U.~R.}\ \bibnamefont
  {Strand}}, \bibinfo {author} {\bibfnamefont {M.}~\bibnamefont {Eckstein}}, \
  and\ \bibinfo {author} {\bibfnamefont {P.}~\bibnamefont {Werner}},\ }\href
  {\doibase 10.1103/PhysRevB.97.235125} {\bibfield  {journal} {\bibinfo
  {journal} {Phys. Rev. B}\ }\textbf {\bibinfo {volume} {97}},\ \bibinfo
  {pages} {235125} (\bibinfo {year} {2018})}\BibitemShut {NoStop}%
\bibitem [{\citenamefont {Tsuji}\ \emph {et~al.}(2016)\citenamefont {Tsuji},
  \citenamefont {Murakami},\ and\ \citenamefont {Aoki}}]{Tsuji2016}%
  \BibitemOpen
  \bibfield  {author} {\bibinfo {author} {\bibfnamefont {N.}~\bibnamefont
  {Tsuji}}, \bibinfo {author} {\bibfnamefont {Y.}~\bibnamefont {Murakami}}, \
  and\ \bibinfo {author} {\bibfnamefont {H.}~\bibnamefont {Aoki}},\ }\href
  {\doibase 10.1103/PhysRevB.94.224519} {\bibfield  {journal} {\bibinfo
  {journal} {Phys. Rev. B}\ }\textbf {\bibinfo {volume} {94}},\ \bibinfo
  {pages} {224519} (\bibinfo {year} {2016})}\BibitemShut {NoStop}%
\bibitem [{\citenamefont {Murakami}\ \emph
  {et~al.}(2017{\natexlab{a}})\citenamefont {Murakami}, \citenamefont {Tsuji},
  \citenamefont {Eckstein},\ and\ \citenamefont {Werner}}]{Murakami2017}%
  \BibitemOpen
  \bibfield  {author} {\bibinfo {author} {\bibfnamefont {Y.}~\bibnamefont
  {Murakami}}, \bibinfo {author} {\bibfnamefont {N.}~\bibnamefont {Tsuji}},
  \bibinfo {author} {\bibfnamefont {M.}~\bibnamefont {Eckstein}}, \ and\
  \bibinfo {author} {\bibfnamefont {P.}~\bibnamefont {Werner}},\ }\href
  {\doibase 10.1103/PhysRevB.96.045125} {\bibfield  {journal} {\bibinfo
  {journal} {Phys. Rev. B}\ }\textbf {\bibinfo {volume} {96}},\ \bibinfo
  {pages} {045125} (\bibinfo {year} {2017}{\natexlab{a}})}\BibitemShut
  {NoStop}%
\bibitem [{\citenamefont {Fransson}\ \emph {et~al.}(2010)\citenamefont
  {Fransson}, \citenamefont {Eriksson},\ and\ \citenamefont
  {Balatsky}}]{Fransson2010}%
  \BibitemOpen
  \bibfield  {author} {\bibinfo {author} {\bibfnamefont {J.}~\bibnamefont
  {Fransson}}, \bibinfo {author} {\bibfnamefont {O.}~\bibnamefont {Eriksson}},
  \ and\ \bibinfo {author} {\bibfnamefont {A.~V.}\ \bibnamefont {Balatsky}},\
  }\href {\doibase 10.1103/PhysRevB.81.115454} {\bibfield  {journal} {\bibinfo
  {journal} {Phys. Rev. B}\ }\textbf {\bibinfo {volume} {81}},\ \bibinfo
  {pages} {115454} (\bibinfo {year} {2010})}\BibitemShut {NoStop}%
\bibitem [{\citenamefont {Genkin}(1997)}]{Genkin1997}%
  \BibitemOpen
  \bibfield  {author} {\bibinfo {author} {\bibfnamefont {G.~M.}\ \bibnamefont
  {Genkin}},\ }\href {\doibase 10.1103/PhysRevB.55.5631} {\bibfield  {journal}
  {\bibinfo  {journal} {Phys. Rev. B}\ }\textbf {\bibinfo {volume} {55}},\
  \bibinfo {pages} {5631} (\bibinfo {year} {1997})}\BibitemShut {NoStop}%
\bibitem [{\citenamefont {Fransson}(2010)}]{Fransson2010b}%
  \BibitemOpen
  \bibfield  {author} {\bibinfo {author} {\bibfnamefont {J.}~\bibnamefont
  {Fransson}},\ }\href {\doibase 10.1103/PhysRevB.82.180411} {\bibfield
  {journal} {\bibinfo  {journal} {Phys. Rev. B}\ }\textbf {\bibinfo {volume}
  {82}},\ \bibinfo {pages} {180411} (\bibinfo {year} {2010})}\BibitemShut
  {NoStop}%
\bibitem [{\citenamefont {Power}\ \emph {et~al.}(2012)\citenamefont {Power},
  \citenamefont {Guimar{\~{a}}es}, \citenamefont {Costa}, \citenamefont
  {Muniz},\ and\ \citenamefont {Ferreira}}]{Power2012}%
  \BibitemOpen
  \bibfield  {author} {\bibinfo {author} {\bibfnamefont {S.~R.}\ \bibnamefont
  {Power}}, \bibinfo {author} {\bibfnamefont {F.~S.~M.}\ \bibnamefont
  {Guimar{\~{a}}es}}, \bibinfo {author} {\bibfnamefont {A.~T.}\ \bibnamefont
  {Costa}}, \bibinfo {author} {\bibfnamefont {R.~B.}\ \bibnamefont {Muniz}}, \
  and\ \bibinfo {author} {\bibfnamefont {M.~S.}\ \bibnamefont {Ferreira}},\
  }\href {\doibase 10.1103/PhysRevB.85.195411} {\bibfield  {journal} {\bibinfo
  {journal} {Phys. Rev. B}\ }\textbf {\bibinfo {volume} {85}},\ \bibinfo
  {pages} {195411} (\bibinfo {year} {2012})}\BibitemShut {NoStop}%
\bibitem [{\citenamefont {Guimar{\~{a}}es}\ \emph {et~al.}(2016)\citenamefont
  {Guimar{\~{a}}es}, \citenamefont {Duffy}, \citenamefont {Costa},
  \citenamefont {Muniz},\ and\ \citenamefont {Ferreira}}]{Guimaraes2016}%
  \BibitemOpen
  \bibfield  {author} {\bibinfo {author} {\bibfnamefont {F.~S.~M.}\
  \bibnamefont {Guimar{\~{a}}es}}, \bibinfo {author} {\bibfnamefont
  {J.}~\bibnamefont {Duffy}}, \bibinfo {author} {\bibfnamefont {A.~T.}\
  \bibnamefont {Costa}}, \bibinfo {author} {\bibfnamefont {R.~B.}\ \bibnamefont
  {Muniz}}, \ and\ \bibinfo {author} {\bibfnamefont {M.~S.}\ \bibnamefont
  {Ferreira}},\ }\href {\doibase 10.1103/PhysRevB.94.235439} {\bibfield
  {journal} {\bibinfo  {journal} {Phys. Rev. B}\ }\textbf {\bibinfo {volume}
  {94}},\ \bibinfo {pages} {235439} (\bibinfo {year} {2016})}\BibitemShut
  {NoStop}%
\bibitem [{\citenamefont {Stephanovich}\ \emph {et~al.}(2017)\citenamefont
  {Stephanovich}, \citenamefont {Dugaev}, \citenamefont {Litvinov},\ and\
  \citenamefont {Berakdar}}]{Stephanovich2017}%
  \BibitemOpen
  \bibfield  {author} {\bibinfo {author} {\bibfnamefont {V.~A.}\ \bibnamefont
  {Stephanovich}}, \bibinfo {author} {\bibfnamefont {V.~K.}\ \bibnamefont
  {Dugaev}}, \bibinfo {author} {\bibfnamefont {V.~I.}\ \bibnamefont
  {Litvinov}}, \ and\ \bibinfo {author} {\bibfnamefont {J.}~\bibnamefont
  {Berakdar}},\ }\href {\doibase 10.1103/PhysRevB.95.045307} {\bibfield
  {journal} {\bibinfo  {journal} {Phys. Rev. B}\ }\textbf {\bibinfo {volume}
  {95}},\ \bibinfo {pages} {045307} (\bibinfo {year} {2017})}\BibitemShut
  {NoStop}%
\bibitem [{\citenamefont {Duan}\ \emph {et~al.}(2018)\citenamefont {Duan},
  \citenamefont {Wang}, \citenamefont {Zheng}, \citenamefont {Wang},
  \citenamefont {Pan},\ and\ \citenamefont {Yang}}]{Duan2018}%
  \BibitemOpen
  \bibfield  {author} {\bibinfo {author} {\bibfnamefont {H.-J.}\ \bibnamefont
  {Duan}}, \bibinfo {author} {\bibfnamefont {C.}~\bibnamefont {Wang}}, \bibinfo
  {author} {\bibfnamefont {S.-H.}\ \bibnamefont {Zheng}}, \bibinfo {author}
  {\bibfnamefont {R.-Q.}\ \bibnamefont {Wang}}, \bibinfo {author}
  {\bibfnamefont {D.-R.}\ \bibnamefont {Pan}}, \ and\ \bibinfo {author}
  {\bibfnamefont {M.}~\bibnamefont {Yang}},\ }\href {\doibase
  10.1038/s41598-018-24567-w} {\bibfield  {journal} {\bibinfo  {journal} {Sci.
  Rep.}\ }\textbf {\bibinfo {volume} {8}},\ \bibinfo {pages} {6185} (\bibinfo
  {year} {2018})}\BibitemShut {NoStop}%
\bibitem [{\citenamefont {Bauer}\ \emph {et~al.}(2015)\citenamefont {Bauer},
  \citenamefont {Babadi},\ and\ \citenamefont {Demler}}]{Bauer2014}%
  \BibitemOpen
  \bibfield  {author} {\bibinfo {author} {\bibfnamefont {J.}~\bibnamefont
  {Bauer}}, \bibinfo {author} {\bibfnamefont {M.}~\bibnamefont {Babadi}}, \
  and\ \bibinfo {author} {\bibfnamefont {E.}~\bibnamefont {Demler}},\ }\href
  {\doibase 10.1103/PhysRevB.92.024305} {\bibfield  {journal} {\bibinfo
  {journal} {Phys. Rev. B}\ }\textbf {\bibinfo {volume} {92}},\ \bibinfo
  {pages} {024305} (\bibinfo {year} {2015})}\BibitemShut {NoStop}%
\bibitem [{\citenamefont {Ribeiro}\ \emph {et~al.}(2015)\citenamefont
  {Ribeiro}, \citenamefont {Zamani},\ and\ \citenamefont
  {Kirchner}}]{Ribeiro2015}%
  \BibitemOpen
  \bibfield  {author} {\bibinfo {author} {\bibfnamefont {P.}~\bibnamefont
  {Ribeiro}}, \bibinfo {author} {\bibfnamefont {F.}~\bibnamefont {Zamani}}, \
  and\ \bibinfo {author} {\bibfnamefont {S.}~\bibnamefont {Kirchner}},\ }\href
  {\doibase 10.1103/PhysRevLett.115.220602} {\bibfield  {journal} {\bibinfo
  {journal} {Phys. Rev. Lett.}\ }\textbf {\bibinfo {volume} {115}},\ \bibinfo
  {pages} {220602} (\bibinfo {year} {2015})}\BibitemShut {NoStop}%
\bibitem [{\citenamefont {Ribeiro}\ \emph {et~al.}(2016)\citenamefont
  {Ribeiro}, \citenamefont {Antipov},\ and\ \citenamefont
  {Rubtsov}}]{Ribeiro2016}%
  \BibitemOpen
  \bibfield  {author} {\bibinfo {author} {\bibfnamefont {P.}~\bibnamefont
  {Ribeiro}}, \bibinfo {author} {\bibfnamefont {A.~E.}\ \bibnamefont
  {Antipov}}, \ and\ \bibinfo {author} {\bibfnamefont {A.~N.}\ \bibnamefont
  {Rubtsov}},\ }\href {\doibase 10.1103/PhysRevB.93.144305} {\bibfield
  {journal} {\bibinfo  {journal} {Phys. Rev. B}\ }\textbf {\bibinfo {volume}
  {93}},\ \bibinfo {pages} {144305} (\bibinfo {year} {2016})}\BibitemShut
  {NoStop}%
\bibitem [{\citenamefont {Ohnuma}\ \emph {et~al.}(2017)\citenamefont {Ohnuma},
  \citenamefont {Matsuo},\ and\ \citenamefont {Maekawa}}]{Ohnuma2017}%
  \BibitemOpen
  \bibfield  {author} {\bibinfo {author} {\bibfnamefont {Y.}~\bibnamefont
  {Ohnuma}}, \bibinfo {author} {\bibfnamefont {M.}~\bibnamefont {Matsuo}}, \
  and\ \bibinfo {author} {\bibfnamefont {S.}~\bibnamefont {Maekawa}},\ }\href
  {\doibase 10.1103/PhysRevB.96.134412} {\bibfield  {journal} {\bibinfo
  {journal} {Phys. Rev. B}\ }\textbf {\bibinfo {volume} {96}},\ \bibinfo
  {pages} {134412} (\bibinfo {year} {2017})}\BibitemShut {NoStop}%
\bibitem [{\citenamefont {Matsuo}\ \emph {et~al.}(2018)\citenamefont {Matsuo},
  \citenamefont {Ohnuma}, \citenamefont {Kato},\ and\ \citenamefont
  {Maekawa}}]{Matsuo2018}%
  \BibitemOpen
  \bibfield  {author} {\bibinfo {author} {\bibfnamefont {M.}~\bibnamefont
  {Matsuo}}, \bibinfo {author} {\bibfnamefont {Y.}~\bibnamefont {Ohnuma}},
  \bibinfo {author} {\bibfnamefont {T.}~\bibnamefont {Kato}}, \ and\ \bibinfo
  {author} {\bibfnamefont {S.}~\bibnamefont {Maekawa}},\ }\href {\doibase
  10.1103/PhysRevLett.120.037201} {\bibfield  {journal} {\bibinfo  {journal}
  {Phys. Rev. Lett.}\ }\textbf {\bibinfo {volume} {120}},\ \bibinfo {pages}
  {037201} (\bibinfo {year} {2018})}\BibitemShut {NoStop}%
\bibitem [{\citenamefont {Wang}\ \emph {et~al.}(2018)\citenamefont {Wang},
  \citenamefont {Chen}, \citenamefont {Moritz},\ and\ \citenamefont
  {Devereaux}}]{Wang2018}%
  \BibitemOpen
  \bibfield  {author} {\bibinfo {author} {\bibfnamefont {Y.}~\bibnamefont
  {Wang}}, \bibinfo {author} {\bibfnamefont {C.-C.}\ \bibnamefont {Chen}},
  \bibinfo {author} {\bibfnamefont {B.}~\bibnamefont {Moritz}}, \ and\ \bibinfo
  {author} {\bibfnamefont {T.~P.}\ \bibnamefont {Devereaux}},\ }\href {\doibase
  10.1103/PhysRevLett.120.246402} {\bibfield  {journal} {\bibinfo  {journal}
  {Phys. Rev. Lett.}\ }\textbf {\bibinfo {volume} {120}},\ \bibinfo {pages}
  {246402} (\bibinfo {year} {2018})}\BibitemShut {NoStop}%
\bibitem [{\citenamefont {Agarwalla}\ \emph {et~al.}(2016)\citenamefont
  {Agarwalla}, \citenamefont {Kulkarni}, \citenamefont {Mukamel},\ and\
  \citenamefont {Segal}}]{Agarwalla2016}%
  \BibitemOpen
  \bibfield  {author} {\bibinfo {author} {\bibfnamefont {B.~K.}\ \bibnamefont
  {Agarwalla}}, \bibinfo {author} {\bibfnamefont {M.}~\bibnamefont {Kulkarni}},
  \bibinfo {author} {\bibfnamefont {S.}~\bibnamefont {Mukamel}}, \ and\
  \bibinfo {author} {\bibfnamefont {D.}~\bibnamefont {Segal}},\ }\href
  {\doibase 10.1103/PhysRevB.94.035434} {\bibfield  {journal} {\bibinfo
  {journal} {Phys. Rev. B}\ }\textbf {\bibinfo {volume} {94}},\ \bibinfo
  {pages} {035434} (\bibinfo {year} {2016})}\BibitemShut {NoStop}%
\bibitem [{\citenamefont {Gole{\v{z}}}\ \emph {et~al.}(2017)\citenamefont
  {Gole{\v{z}}}, \citenamefont {Boehnke}, \citenamefont {Strand}, \citenamefont
  {Eckstein},\ and\ \citenamefont {Werner}}]{Golez2017}%
  \BibitemOpen
  \bibfield  {author} {\bibinfo {author} {\bibfnamefont {D.}~\bibnamefont
  {Gole{\v{z}}}}, \bibinfo {author} {\bibfnamefont {L.}~\bibnamefont
  {Boehnke}}, \bibinfo {author} {\bibfnamefont {H.~U.~R.}\ \bibnamefont
  {Strand}}, \bibinfo {author} {\bibfnamefont {M.}~\bibnamefont {Eckstein}}, \
  and\ \bibinfo {author} {\bibfnamefont {P.}~\bibnamefont {Werner}},\ }\href
  {\doibase 10.1103/PhysRevLett.118.246402} {\bibfield  {journal} {\bibinfo
  {journal} {Phys. Rev. Lett.}\ }\textbf {\bibinfo {volume} {118}},\ \bibinfo
  {pages} {246402} (\bibinfo {year} {2017})}\BibitemShut {NoStop}%
\bibitem [{\citenamefont {Ribeiro}(2017)}]{Ribeiro2017}%
  \BibitemOpen
  \bibfield  {author} {\bibinfo {author} {\bibfnamefont {P.}~\bibnamefont
  {Ribeiro}},\ }\href {\doibase 10.1103/PhysRevB.96.054302} {\bibfield
  {journal} {\bibinfo  {journal} {Phys. Rev. B}\ }\textbf {\bibinfo {volume}
  {96}},\ \bibinfo {pages} {054302} (\bibinfo {year} {2017})}\BibitemShut
  {NoStop}%
\bibitem [{\citenamefont {B{\"{u}}nemann}\ and\ \citenamefont
  {Seibold}(2017)}]{Bunemann2017}%
  \BibitemOpen
  \bibfield  {author} {\bibinfo {author} {\bibfnamefont {J.}~\bibnamefont
  {B{\"{u}}nemann}}\ and\ \bibinfo {author} {\bibfnamefont {G.}~\bibnamefont
  {Seibold}},\ }\href {\doibase 10.1103/PhysRevB.96.245139} {\bibfield
  {journal} {\bibinfo  {journal} {Phys. Rev. B}\ }\textbf {\bibinfo {volume}
  {96}},\ \bibinfo {pages} {245139} (\bibinfo {year} {2017})}\BibitemShut
  {NoStop}%
\bibitem [{\citenamefont {Murakami}\ \emph
  {et~al.}(2016{\natexlab{a}})\citenamefont {Murakami}, \citenamefont {Werner},
  \citenamefont {Tsuji},\ and\ \citenamefont {Aoki}}]{Murakami2016}%
  \BibitemOpen
  \bibfield  {author} {\bibinfo {author} {\bibfnamefont {Y.}~\bibnamefont
  {Murakami}}, \bibinfo {author} {\bibfnamefont {P.}~\bibnamefont {Werner}},
  \bibinfo {author} {\bibfnamefont {N.}~\bibnamefont {Tsuji}}, \ and\ \bibinfo
  {author} {\bibfnamefont {H.}~\bibnamefont {Aoki}},\ }\href {\doibase
  10.1103/PhysRevB.93.094509} {\bibfield  {journal} {\bibinfo  {journal} {Phys.
  Rev. B}\ }\textbf {\bibinfo {volume} {93}},\ \bibinfo {pages} {094509}
  (\bibinfo {year} {2016}{\natexlab{a}})}\BibitemShut {NoStop}%
\bibitem [{\citenamefont {Murakami}\ \emph
  {et~al.}(2016{\natexlab{b}})\citenamefont {Murakami}, \citenamefont {Werner},
  \citenamefont {Tsuji},\ and\ \citenamefont {Aoki}}]{Murakami2016b}%
  \BibitemOpen
  \bibfield  {author} {\bibinfo {author} {\bibfnamefont {Y.}~\bibnamefont
  {Murakami}}, \bibinfo {author} {\bibfnamefont {P.}~\bibnamefont {Werner}},
  \bibinfo {author} {\bibfnamefont {N.}~\bibnamefont {Tsuji}}, \ and\ \bibinfo
  {author} {\bibfnamefont {H.}~\bibnamefont {Aoki}},\ }\href {\doibase
  10.1103/PhysRevB.94.115126} {\bibfield  {journal} {\bibinfo  {journal} {Phys.
  Rev. B}\ }\textbf {\bibinfo {volume} {94}},\ \bibinfo {pages} {115126}
  (\bibinfo {year} {2016}{\natexlab{b}})}\BibitemShut {NoStop}%
\bibitem [{\citenamefont {Fischer}\ \emph {et~al.}(2018)\citenamefont
  {Fischer}, \citenamefont {Hecker}, \citenamefont {Hoyer},\ and\ \citenamefont
  {Schmalian}}]{Fischer2018}%
  \BibitemOpen
  \bibfield  {author} {\bibinfo {author} {\bibfnamefont {S.}~\bibnamefont
  {Fischer}}, \bibinfo {author} {\bibfnamefont {M.}~\bibnamefont {Hecker}},
  \bibinfo {author} {\bibfnamefont {M.}~\bibnamefont {Hoyer}}, \ and\ \bibinfo
  {author} {\bibfnamefont {J.}~\bibnamefont {Schmalian}},\ }\href {\doibase
  10.1103/PhysRevB.97.054510} {\bibfield  {journal} {\bibinfo  {journal} {Phys.
  Rev. B}\ }\textbf {\bibinfo {volume} {97}},\ \bibinfo {pages} {054510}
  (\bibinfo {year} {2018})}\BibitemShut {NoStop}%
\bibitem [{\citenamefont {Ido}\ \emph {et~al.}(2017)\citenamefont {Ido},
  \citenamefont {Ohgoe},\ and\ \citenamefont {Imada}}]{Ido2017}%
  \BibitemOpen
  \bibfield  {author} {\bibinfo {author} {\bibfnamefont {K.}~\bibnamefont
  {Ido}}, \bibinfo {author} {\bibfnamefont {T.}~\bibnamefont {Ohgoe}}, \ and\
  \bibinfo {author} {\bibfnamefont {M.}~\bibnamefont {Imada}},\ }\href
  {\doibase 10.1126/sciadv.1700718} {\bibfield  {journal} {\bibinfo  {journal}
  {Sci. Adv.}\ }\textbf {\bibinfo {volume} {3}},\ \bibinfo {pages} {e1700718}
  (\bibinfo {year} {2017})}\BibitemShut {NoStop}%
\bibitem [{\citenamefont {Murakami}\ \emph
  {et~al.}(2017{\natexlab{b}})\citenamefont {Murakami}, \citenamefont
  {Gole{\v{z}}}, \citenamefont {Eckstein},\ and\ \citenamefont
  {Werner}}]{Murakami2017b}%
  \BibitemOpen
  \bibfield  {author} {\bibinfo {author} {\bibfnamefont {Y.}~\bibnamefont
  {Murakami}}, \bibinfo {author} {\bibfnamefont {D.}~\bibnamefont
  {Gole{\v{z}}}}, \bibinfo {author} {\bibfnamefont {M.}~\bibnamefont
  {Eckstein}}, \ and\ \bibinfo {author} {\bibfnamefont {P.}~\bibnamefont
  {Werner}},\ }\href {\doibase 10.1103/PhysRevLett.119.247601} {\bibfield
  {journal} {\bibinfo  {journal} {Phys. Rev. Lett.}\ }\textbf {\bibinfo
  {volume} {119}},\ \bibinfo {pages} {247601} (\bibinfo {year}
  {2017}{\natexlab{b}})}\BibitemShut {NoStop}%
\bibitem [{\citenamefont {Tsuji}\ \emph {et~al.}(2008)\citenamefont {Tsuji},
  \citenamefont {Oka},\ and\ \citenamefont {Aoki}}]{Tsuji2008}%
  \BibitemOpen
  \bibfield  {author} {\bibinfo {author} {\bibfnamefont {N.}~\bibnamefont
  {Tsuji}}, \bibinfo {author} {\bibfnamefont {T.}~\bibnamefont {Oka}}, \ and\
  \bibinfo {author} {\bibfnamefont {H.}~\bibnamefont {Aoki}},\ }\href {\doibase
  10.1103/PhysRevB.78.235124} {\bibfield  {journal} {\bibinfo  {journal} {Phys.
  Rev. B}\ }\textbf {\bibinfo {volume} {78}},\ \bibinfo {pages} {235124}
  (\bibinfo {year} {2008})}\BibitemShut {NoStop}%
\bibitem [{Note1()}]{Note1}%
  \BibitemOpen
  \bibinfo {note} {In this paper, we call the limits of $\protect \mathit
  {\Omega }\rightarrow 0$ and $\omega \rightarrow 0$ the low-frequency limit
  and the static limit, respectively}\BibitemShut {NoStop}%
\bibitem [{\citenamefont {Ruderman}\ and\ \citenamefont
  {Kittel}(1954)}]{Ruderman1954}%
  \BibitemOpen
  \bibfield  {author} {\bibinfo {author} {\bibfnamefont {M.~A.}\ \bibnamefont
  {Ruderman}}\ and\ \bibinfo {author} {\bibfnamefont {C.}~\bibnamefont
  {Kittel}},\ }\href {\doibase 10.1103/PhysRev.96.99} {\bibfield  {journal}
  {\bibinfo  {journal} {Phys. Rev.}\ }\textbf {\bibinfo {volume} {96}},\
  \bibinfo {pages} {99} (\bibinfo {year} {1954})}\BibitemShut {NoStop}%
\bibitem [{\citenamefont {Kasuya}(1956)}]{Kasuya1956}%
  \BibitemOpen
  \bibfield  {author} {\bibinfo {author} {\bibfnamefont {T.}~\bibnamefont
  {Kasuya}},\ }\href {\doibase 10.1143/PTP.16.45} {\bibfield  {journal}
  {\bibinfo  {journal} {Prog. Theor. Phys.}\ }\textbf {\bibinfo {volume}
  {16}},\ \bibinfo {pages} {45} (\bibinfo {year} {1956})}\BibitemShut {NoStop}%
\bibitem [{\citenamefont {Yosida}(1957)}]{Yosida1957}%
  \BibitemOpen
  \bibfield  {author} {\bibinfo {author} {\bibfnamefont {K.}~\bibnamefont
  {Yosida}},\ }\href {\doibase 10.1103/PhysRev.106.893} {\bibfield  {journal}
  {\bibinfo  {journal} {Phys. Rev.}\ }\textbf {\bibinfo {volume} {106}},\
  \bibinfo {pages} {893} (\bibinfo {year} {1957})}\BibitemShut {NoStop}%
\bibitem [{\citenamefont {Dunlap}\ and\ \citenamefont
  {Kenkre}(1986)}]{Dunlap1986}%
  \BibitemOpen
  \bibfield  {author} {\bibinfo {author} {\bibfnamefont {D.~H.}\ \bibnamefont
  {Dunlap}}\ and\ \bibinfo {author} {\bibfnamefont {V.~M.}\ \bibnamefont
  {Kenkre}},\ }\href {\doibase 10.1103/PhysRevB.34.3625} {\bibfield  {journal}
  {\bibinfo  {journal} {Phys. Rev. B}\ }\textbf {\bibinfo {volume} {34}},\
  \bibinfo {pages} {3625} (\bibinfo {year} {1986})}\BibitemShut {NoStop}%
\bibitem [{\citenamefont {Holthaus}(1992)}]{Holthaus1992}%
  \BibitemOpen
  \bibfield  {author} {\bibinfo {author} {\bibfnamefont {M.}~\bibnamefont
  {Holthaus}},\ }\href {\doibase 10.1103/PhysRevLett.69.351} {\bibfield
  {journal} {\bibinfo  {journal} {Phys. Rev. Lett.}\ }\textbf {\bibinfo
  {volume} {69}},\ \bibinfo {pages} {351} (\bibinfo {year} {1992})}\BibitemShut
  {NoStop}%
\bibitem [{\citenamefont {Grossmann}\ \emph {et~al.}(1991)\citenamefont
  {Grossmann}, \citenamefont {Dittrich}, \citenamefont {Jung},\ and\
  \citenamefont {H{\"{a}}nggi}}]{Grossmann1991}%
  \BibitemOpen
  \bibfield  {author} {\bibinfo {author} {\bibfnamefont {F.}~\bibnamefont
  {Grossmann}}, \bibinfo {author} {\bibfnamefont {T.}~\bibnamefont {Dittrich}},
  \bibinfo {author} {\bibfnamefont {P.}~\bibnamefont {Jung}}, \ and\ \bibinfo
  {author} {\bibfnamefont {P.}~\bibnamefont {H{\"{a}}nggi}},\ }\href {\doibase
  10.1103/PhysRevLett.67.516} {\bibfield  {journal} {\bibinfo  {journal} {Phys.
  Rev. Lett.}\ }\textbf {\bibinfo {volume} {67}},\ \bibinfo {pages} {516}
  (\bibinfo {year} {1991})}\BibitemShut {NoStop}%
\bibitem [{\citenamefont {Kayanuma}\ and\ \citenamefont
  {Saito}(2008)}]{Kayanuma2008}%
  \BibitemOpen
  \bibfield  {author} {\bibinfo {author} {\bibfnamefont {Y.}~\bibnamefont
  {Kayanuma}}\ and\ \bibinfo {author} {\bibfnamefont {K.}~\bibnamefont
  {Saito}},\ }\href {\doibase 10.1103/PhysRevA.77.010101} {\bibfield  {journal}
  {\bibinfo  {journal} {Phys. Rev. A}\ }\textbf {\bibinfo {volume} {77}},\
  \bibinfo {pages} {010101} (\bibinfo {year} {2008})}\BibitemShut {NoStop}%
\bibitem [{Note2()}]{Note2}%
  \BibitemOpen
  \bibinfo {note} {There is a deviation between the numerically exact
  susceptibility and the approximated one at $q=0$. This is because the former
  is computed via the fast Fourier transformation of Eq.~\protect \textup
  {\hbox {\mathsurround \z@ \protect \normalfont (\ignorespaces \ref
  {eq:susceptibility_floquet}\unskip \@@italiccorr )}}, whereas the latter is
  directly evaluated from Eqs.~\protect \textup {\hbox {\mathsurround \z@
  \protect \normalfont (\ignorespaces \ref {eq:expansion_chi_base}\unskip
  \@@italiccorr )}}--\protect \textup {\hbox {\mathsurround \z@ \protect
  \normalfont (\ignorespaces \ref {eq:expansion_chi_inter}\unskip \@@italiccorr
  )}}, which vanish since all of the numerators are zero for
  $q=0$.}\BibitemShut {Stop}%
\end{thebibliography}%
\end{document}